\documentclass[preprint,showpacs,preprintnumbers,amsmath,amssymb]{revtex4}
\usepackage{graphicx}
\usepackage{dcolumn}
\usepackage{bm}

\begin{document}
\preprint{ECT*-05-24}
\title{Non perturbative renormalisation group and momentum dependence of $n$-point functions (I)}
\author{Jean-Paul Blaizot}
  \thanks{CNRS fellow} \email{blaizot@ect.it}
  \affiliation{ECT*, Villa Tambosi, strada delle Tabarelle 286, 38050 Villazzano (TN), Italy}
  \author{Ram\'on M\'endez-Galain}
  \email{mendezg@fing.edu.uy}
  \affiliation{Instituto de F\'{\i}sica, Facultad de Ingenier\'{\i}a, J.H.y Reissig 565, 11000
  Montevideo, Uruguay}

\author{Nicol\'as Wschebor}
  \email{nicws@fing.edu.uy}
  \affiliation{
  Instituto de F\'{\i}sica, Facultad de Ingenier\'{\i}a, J.H.y Reissig 565, 11000
  Montevideo, Uruguay}

  \date{\today}
\begin{abstract}

We present an approximation scheme to solve the Non Perturbative
Renormalization Group equations and obtain  the full
momentum dependence of the $n$-point functions. It is based on an
iterative procedure where, in a first step, an initial ansatz for the $n$-point functions is
constructed by solving approximate  flow equations derived from well
motivated approximations. These approximations exploit the derivative
expansion and the 
decoupling of high momentum modes. The method is applied to the O($N$) model. In leading order, the self energy is already accurate both in the
perturbative and the scaling regimes. A stringent test is provided by  the
calculation of the shift $\Delta T_c$ in the transition
temperature of the weakly repulsive Bose gas, a quantity which is
particularly sensitive to all momentum scales. The leading order
result is in agreement with lattice calculations,
  albeit with a theoretical uncertainty of about 25\%.
  \end{abstract}

\maketitle


\def\bfphi{\mbox{\boldmath$\phi$}}
\def\bfvarphi{\mbox{\boldmath$\varphi$}}
\def\bfgamma{\mbox{\boldmath$\gamma$}}
\def\bfalpha{\mbox{\boldmath$\alpha$}}
\def\bftau{\mbox{\boldmath$\tau$}}
\def\bfnabla{\mbox{\boldmath$\nabla$}}
\def\bfsigma{\mbox{\boldmath$\sigma$}}
\def\bfpi{\mbox{\boldmath$\pi$}}

\newcommand \beq{\begin{eqnarray}}
\newcommand \eeq{\end{eqnarray}}

\newcommand \ga{\raisebox{-.5ex}{$\stackrel{>}{\sim}$}}
\newcommand \la{\raisebox{-.5ex}{$\stackrel{<}{\sim}$}}

\def\psib{\psi}
\def\phib{\phi}
\def\r{{\rm r}}
\def\d{{\rm d}}

\def \e {\mbox{e}}

\input epsf


\def\square{\hbox{{$\sqcup$}\llap{$\sqcap$}}}
\def\grad{\nabla}
\def\del{\partial}

\def\frac#1#2{{#1 \over #2}}
\def\smallfrac#1#2{{\scriptstyle {#1 \over #2}}}
\def\half{\ifinner {\scriptstyle {1 \over 2}}
   \else {1 \over 2} \fi}


\def\bra#1{\langle#1\vert}
\def\ket#1{\vert#1\rangle}


\def\simge{\mathrel{%
   \rlap{\raise 0.511ex \hbox{$>$}}{\lower 0.511ex \hbox{$\sim$}}}}
\def\simle{\mathrel{
   \rlap{\raise 0.511ex \hbox{$<$}}{\lower 0.511ex \hbox{$\sim$}}}}


\def\buildchar#1#2#3{{\null\!
   \mathop#1\limits^{#2}_{#3}
   \!\null}}
\def\overcirc#1{\buildchar{#1}{\circ}{}}


\def\slashchar#1{\setbox0=\hbox{$#1$}
   \dimen0=\wd0
   \setbox1=\hbox{/} \dimen1=\wd1
   \ifdim\dimen0>\dimen1
      \rlap{\hbox to \dimen0{\hfil/\hfil}}
      #1
   \else
      \rlap{\hbox to \dimen1{\hfil$#1$\hfil}}
      /
   \fi}


\def\real{\mathop{\rm Re}\nolimits}     
\def\imag{\mathop{\rm Im}\nolimits}     

\def\tr{\mathop{\rm tr}\nolimits}       
\def\Tr{\mathop{\rm Tr}\nolimits}       
\def\Det{\mathop{\rm Det}\nolimits}     

\def\mod{\mathop{\rm mod}\nolimits}     
\def\wrt{\mathop{\rm wrt}\nolimits}     


\def\TeV{{\rm TeV}}                     
\def\GeV{{\rm GeV}}                     
\def\MeV{{\rm MeV}}                     
\def\KeV{{\rm KeV}}                     
\def\eV{{\rm eV}}                       

\def\mb{{\rm mb}}                       
\def\mub{\hbox{$\mu$b}}                 
\def\nb{{\rm nb}}                       
\def\pb{{\rm pb}}                       

%
%

\def\picture #1 by #2 (#3){
  \vbox to #2{
    \hrule width #1 height 0pt depth 0pt
    \vfill
    \special{picture #3} 
    }
  }

\def\scaledpicture #1 by #2 (#3 scaled #4){{
  \dimen0=#1 \dimen1=#2
  \divide\dimen0 by 1000 \multiply\dimen0 by #4
  \divide\dimen1 by 1000 \multiply\dimen1 by #4
  \picture \dimen0 by \dimen1 (#3 scaled #4)}
  }

\def\centerpicture #1 by #2 (#3 scaled #4){
   \dimen0=#1 \dimen1=#2
    \divide\dimen0 by 1000 \multiply\dimen0 by #4
    \divide\dimen1 by 1000 \multiply\dimen1 by #4
         \noindent
         \vbox{
            \hspace*{\fill}
            \picture \dimen0 by \dimen1 (#3 scaled #4)
            \hspace*{\fill}
            \vfill}}


\def\figfermass{\centerpicture 122.4mm by 32.46mm
 (fermass scaled 750)}

%

\section{Introduction}

The need for reliable and efficient  non-perturbative calculation methods is felt, in various forms, in
nearly all fields of physics: in nuclear and particle physics (to deal with the
infrared sector of quantum chromodynamics and the associated  phenomena of
color confinement and chiral symmetry breaking), in condensed
matter and statistical physics of systems in or out of equilibrium
(phase transitions and critical phenomena, disorder systems,
strongly correlated electrons), to quote just but a few general
examples. In many of these cases,  the absence of a small
parameter prevents one to build a solution in terms of a
systematic expansion. In order to treat such problems, what one needs is a non-perturbative method that allows the calculation of correlation functions for arbitrary values of the external momenta, from which most physical quantities can be deduced. 

Among the non perturbative methods that have been developed along the years,  the non perturbative renormalization group
(NPRG) \cite{Wetterich93,Ellwanger93,Tetradis94,Morris94,Morris94c} stands out as a
very promising tool, suggesting
 new approximation schemes which are not easily formulated in other,
more conventional, approaches in field theory or many body
physics. The NPRG has been applied successfully to a variety of
physical problems, in condensed matter, particle or nuclear
physics (for  reviews, see e.g.
\cite{Bagnuls:2000ae,Berges02,Canet04}). In most of these problems
however, the focus is on long wavelength modes  and  the solution
of the NPRG equations involves generally a derivative expansion
which only allows for the determination of the $n$-point functions
and their derivatives at small external momenta (vanishing momenta
in the case of critical phenomena). In many situations, this is
not enough: a full knowledge of the momentum dependence of the
correlation functions is needed to calculate the quantities of
physical interest (e.g. to get the spectrum of excitations, the
shape of the Fermi surface, the scattering matrix, etc.).

 The NPRG presents itself as an infinite hierarchy of equations relating sequentially the various $n$-point functions.
To our knowledge, most efforts to solve this hierachy, aside from the derivative expansion alluded to above, 
have been based on various forms
 of the early proposal by Weinberg \cite{weinberg73}, that is they involve some truncation of  the infinite tower of flow equations for the $n$-point 
functions, ignoring higher order vertices, or 
possibly using various ansatzs for some of them  \cite{truncation,Ellwanger94}. 
This leads to approximations similar to those
used when solving the hierarchy of Schwinger-Dyson equations \cite{alkofer}.
However, despite the fact that very encouraging results have been obtained 
  this approximation scheme presents convergence dificulties  \cite{convergence}.

The goal of this paper is then to  present a method for solving
the NPRG equations that keeps the contribution of all the  vertices present in the flow equations.  This is achieved by exploiting specific properties of the NPRG. The method allows one to get, in a relatively simple way, the
full momentum dependence of the $n$-point functions. It involves iterations that starts with an initial guess for the $n$-point functions. That initial guess is then injected in the flow equations which are integrated in order to obtain a leading order expression for the $n$-point functions. And so on. Clearly, each new iteration involves more $n$-point functions, and the scheme may become rapidly prohibitively complicated. It is therefore crucial  that the starting point of the iterations, that is, the initial ansatz for the $n$-point functions, be  as close as possible to the exact solution, in order to get a good approximation with a minimum number of iterations. The construction of this initial ansatz is therefore  the central part of the method.

To derive this initial ansatz we shall first simplify the flow equations using well motivated approximations.  We shall exploit a modified  derivative expansion in its leading order and the decoupling of high momentum modes  in the flow equations in order to simplify the momentum dependence of the vertices that govern the flow. The resulting approximate equations are then solved explicitly.

The particular class of problems that we are concerned
with can be formulated in terms of a field theory, and as a generic case, we shall 
consider here a scalar $\phib^4$
theory in $d$ dimension with $O(N)$ symmetry:
\beq
\label{classicalaction} S = \int {\rm d}^{d}x\, \left\lbrace{ 1
\over 2} \left[ \nabla \varphi (x) \right]^2+{1 \over 2}r
\varphi^2 (x)+{u \over 4!} \left[ \varphi^2(x) \right]^2
\right\rbrace,\eeq 
 where the
field $\varphi(x)$ has $N$ real components $\varphi_i(x)$, with
$i=1,\cdots,N$. 
We emphasize however that  most of the arguments presented in this paper have a wider range of
applicability.

In this paper, we  shall apply the method to   the calculation of the self-energy  at criticality and at zero external field, in leading order and in $d=3$. This involves getting the initial ansatz for  both the propagator and the 4-point function. Constructing this initial ansatz is  the main task carried out in the present paper.  It is presented in sect.~\ref{Approximation_scheme}, together with a more detailed description of the approximation scheme.  First, in sect.~\ref{NPRGgral}, we review basic features of the NPRG, and illustrate various strategies that have been used to obtain solutions of the flow equations. These will provide the necessary background to motivate  the approximation scheme presented in sect.~\ref{Approximation_scheme}, as well as the approximations involved in the construction of the initial ansatz for the 4-point function. The reader familiar with the NPRG may skip this section and go directly to sect.~\ref{Approximation_scheme}. The results for the self-energy are presented in sect.~\ref{sec:LO}. 

The self-energy thus obtained has the correct behavior at all momenta. It agrees with perturbation theory in the ultraviolet and it presents the expected power-law behavior in the infrared. As a
benchmark for our approximation scheme we shall use the  shift
$\Delta T_c$ of the transition temperature of a weakly interacting
Bose gas \cite{club,bigbec} (see also \cite{Andersen:2003qj} for a recent review on the theory of the weakly interacting Bose gas). As we shall recall later, the precise
evaluation of $\Delta T_c$ requires an accurate knowledge of a
2-point function at all momentum scales, and it constitutes
therefore a very stringent test of any method aiming at getting
the full momentum dependence of $n$-point functions.   As shown in
Ref.~\cite{club}, the calculation  of $\Delta T_c$ reduces to that of the change $\Delta\langle
\varphi^2\rangle$ of the magnitude of the fluctuations of the field described by the
action (\ref{classicalaction}), for $d=3$ and $N=2$ \cite{BigN}. This calculation can be done immediately once the self-energy is known. It is presented in sect.~\ref{calcul_de_c} together with a comparison with estimates of this quantity using different techniques, for instance lattice calculations \cite{latt2,latt1}. 

In a companion paper \cite{alpha2},  we extend  the method described here  to the next-to-leading order calculation of the self-energy (which involves the leading order calculation of the 4-point function). Some of the  results of this study have already been presented in ref.~\cite{Blaizot:2004qa}. However, since the publication of ref.~\cite{Blaizot:2004qa}, we have been able to improve the accuracy of the  leading order  calculation of the 4-point function, which yields a considerable improvement of the next-to-leading order self-energy;  the final results that we obtain for $\Delta T_c$ are in excellent agreement with the lattice calculations, with a much reduced  theoretical uncertainty as compared with the estimates presented in the present paper \cite{alpha2}. Further progress has been achieved in an effort to get rid of  some of the approximations used in the present work, and which contributes to the theoretical uncertainty in the predictions. A possible strategy to do so has been presented in \cite{PLB}, and first results concerning its numerical implementation will be presented shortly \cite{BMWn}.

 \section{Some features of the NPRG equations}
\label{NPRGgral}
\subsection{Generalities}

\label{presentation}

The NPRG   allows the construction of a set of
effective actions $\Gamma_\kappa[\phi]$ which interpolate between the classical action $S$ and the full effective action $\Gamma[\phi]$: In
$\Gamma_\kappa[\phi]$   the magnitude of long wavelength
fluctuations of the field is controlled by an infrared regulator
depending on a continuous parameter $\kappa$ which has the dimension of a momentum.  The full effective action is obtained for the value $\kappa=0$, the situation with no infrared cut-off and where therefore  all fluctuations are taken into account. In the other limit, corresponding to a value of $\kappa$ of the order of a microscopic scale $\Lambda$ at  which fluctuations are suppressed, $\Gamma_{\kappa=\Lambda}[\phi]$ reduces to the classical action \footnote{Note that depending on the choice of the regulator, not all fluctuations may be suppressed when $\kappa=\Lambda$. However, for renormalisable theories, and if $\Lambda$ is large enough, the effects of these remnant fluctuations can be absorbed into a redefinition of the parameters of the classical action.}.

In practice the control of the magnitude of the fluctuations
is implemented by adding to the classical  action (\ref{classicalaction})
the regulator
 \beq\label{regulaction}
  \Delta S_\kappa[\varphi] =\frac{1}{2} \int \frac{{\rm d}^dq}{(2\pi)^d}
\, \varphi_i(q)\,R_\kappa(q) \,\varphi_i(-q),
  \eeq\normalsize
where $R_\kappa$ denotes a family of ``cut-off functions''
depending  on  $\kappa$.  The role of $\Delta S_\kappa$ is to
suppress the fluctuations with momenta $q\simle \kappa$, while
leaving unaffected those with $q\simge \kappa$. Thus,
typically, $ R_{\kappa}(q)\to\kappa^2$ when $ q \ll \kappa$, and
$R_{\kappa}(q)\to 0$
 when $ q\simge \kappa$.
There is a large freedom in the choice of $R_\kappa(q)$,
abundantly discussed in the literature
\cite{Ball95,Comellas98,Litim,Canet02}. The choice of the cut-off function matters when approximations are done, as is the case in all situations of practical interest. We have used in this work the cut-off
function proposed in \cite{Litim}:
\beq\label{reg-litim}
R_\kappa(q^2)\propto (\kappa^2-q^2) \theta(\kappa^2-q^2).
\eeq
This regulator allows many calculations to be done analytically. It is known to work well with the derivative expansion in leading order, which we shall use in this work.

For each value of $\kappa$, one defines the generating functional of
connected Green's functions \beq W_\kappa[J]= \ln \int D\varphi
\hspace{.1cm} \exp\left\{  {-S[\varphi]-\Delta S_\kappa[\varphi]+\int
d^dx \varphi(x)J(x)}\right\} . \eeq
We have, for instance, \beq
\label{Legendre1} \phi_{\kappa,J}(x)\equiv\left\langle \varphi(x)
\right\rangle_{\kappa,J} =\frac{\delta W_\kappa}{\delta J(x)}.
\eeq The Feynman diagrams contributing to $W_\kappa$ are those of
ordinary perturbation theory, except that the propagators contain
the infrared regulator.  We also define the effective action,
through a modified Legendre transform that includes the  explicit subtraction of $\Delta S_\kappa$:  \beq
\Gamma_\kappa[\phi]=-W_\kappa[J_{\phi}]+\int d^dx \hspace{.1cm}
\phi(x) J_{\phi}(x)-\Delta S_\kappa[\phi], \eeq where $J_\phi$ is
obtained by inverting eq.~(\ref{Legendre1}). Note that, in this
inversion, $\phi$ is considered as a given variable, so that
$J_\phi$ becomes  implicitly dependent on $\kappa$.

\begin{figure}[t!]
\begin{center}
\includegraphics*[scale=0.8,angle=0]{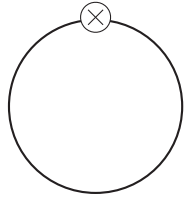}
\end{center}
\caption{ Diagrammatic illustration of the r.h.s. of the  flow equation
of the effective action, eq.~(\ref{mastereq}).
The crossed circle represents
an insertion of $\del_\kappa R_\kappa$, and the thick line a full
propagator in an arbitrary background field.\label{fig:mastereq}}
\end{figure}

One can write an exact flow equation for $\Gamma_\kappa[\phi]$  which
gives its variation as a function of $\kappa$, at fixed $\phi$.
It reads  \cite{Wetterich93,Ellwanger93,Tetradis94,Morris94,Morris94c}:
\beq\label{mastereq}
\partial_\kappa \Gamma_\kappa[\phi]=\frac{1}{2}{\rm tr}\int \frac{d^dq}{(2\pi)^d}
\,\partial_\kappa R_\kappa(q^2)
\left[\Gamma_\kappa^{(2)}+R_\kappa\right]^{-1}_{q,-q},
\eeq
where $\Gamma_\kappa^{(2)}$ is the second derivative of $\Gamma_\kappa$
with respect to $\phi$, and the trace ${\rm tr}$ runs over the O($N$) indices. Eq.~(\ref{mastereq}) is
the master equation of the NPRG. Its right hand side has the
structure of a one loop integral, with one insertion of
$\partial_\kappa R_\kappa(q^2)$ (see fig.~\ref{fig:mastereq}).
The solution of eq.~(\ref{mastereq}) yields the effective action $\Gamma[\phi]=\Gamma_{\kappa=0}[\phi]$ starting with the initial condition $\Gamma_{\kappa=\Lambda}[\phi]=S[\phi]$ (see e.g. \cite{Berges02}).

\begin{figure}[t!]
\begin{center}
\includegraphics*[scale=0.8,angle=0]{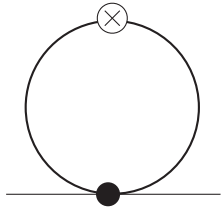}
\end{center}
\caption{\label{fig:dGamma2} Diagrammatic illustration of the r.h.s. of  the flow equation for the  2-point function, eq.~(\ref{eq:dGamma2}). The black dot
denotes the four-point function and the thick line the full propagator $G$. The circled cross represents the insertion of
$\partial_\kappa R_\kappa$.}
\end{figure}

\begin{figure}[t!]
\begin{center}
\includegraphics*[scale=0.8,angle=0]{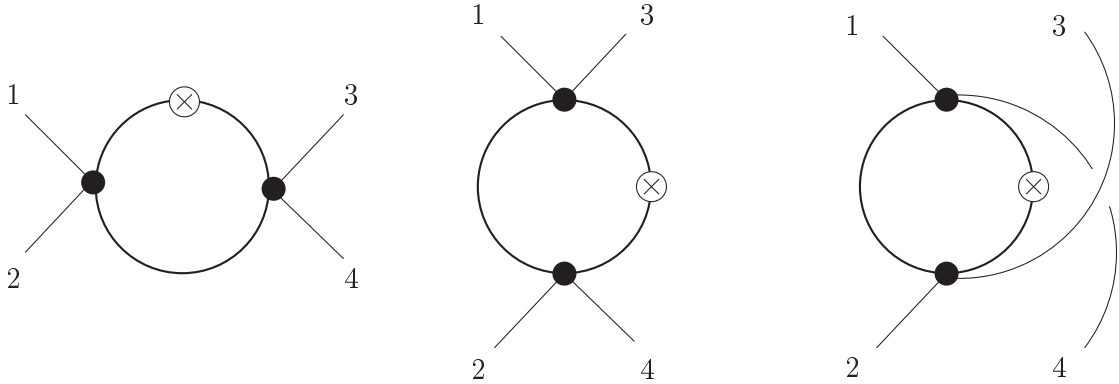}
\end{center}
\caption{\label{fig:dGamma4} Diagrammatic illustration of the r.h.s. of the flow
equation for the 4-point function, eq.~(\ref{4point}): contribution of the 4-point functions (represented by black disks) in the three channels $s$, $t$ and $u$, from left to right. The crossed circle represents
an insertion of $\del_\kappa R_\kappa$, and the thick line a full
propagator.\label{fig:4point}}
\end{figure}

\begin{figure}[t!]
\begin{center}
\includegraphics*[scale=0.8,angle=0]{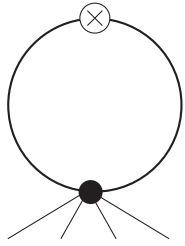}
\end{center}
\caption{\label{fig:dGamma4_6} Diagrammatic illustration of the r.h.s. of the flow
equation for the 4-point function, eq.~(\ref{4point}):  contribution of the 6-point function $\Gamma^{(6)}$ (represented by a black disk). The crossed circle represents
an insertion of $\del_\kappa R_\kappa$, and the thick line a full
propagator.\label{fig:6point}}
\end{figure}

As well known (see e.g. \cite{Zinn-Justin:2002ru}), the effective action
$\Gamma[\phi]$ is the generating functional of the one-particle
irreducible $n$-point functions. This property extends trivially
to $\Gamma_\kappa[\phi]$. Since we shall be concerned only with
$n$-point functions for constant (eventually vanishing) external fields   we exploit
translational invariance to define reduced
$n$-point functions $\Gamma^{(n)}(\kappa;p_1,\dots,p_n)$ as follows :
\beq\label{rednptfcts}
(2\pi)^d \delta^{(d)}(p_1+\cdots +p_n)\,\Gamma^{(n)}(\kappa;p_1,\dots,p_n)\qquad\qquad\qquad\qquad\qquad\qquad\qquad\nonumber\\
=\int d^dx_1\dots\int
d^dx_{n} \,{\rm e}^{i\sum_{j=1}^n
p_j\cdot x_j}\left.\frac{\delta^n\Gamma_\kappa[\phi]}{\delta\phi(x_1)
\dots \delta\phi(x_n)}\right|_{\phi={\rm cst}}.
\eeq
By deriving eq.~(\ref{mastereq}) with respect to $\phi$, and then
letting the field be zero, one gets the flow equations for all
$n$-point functions in a  vanishing background field $\phi$.  For
example, the equation for the 2-point function reads:
\begin{equation}\label{eq:dGamma2}
 \hspace{-1cm}\partial_\kappa
\Gamma^{(2)}_{12}(\kappa;p)\equiv \delta_{12}\del_\kappa \Sigma(\kappa; p)= -\frac{1}{2}\int
\frac{d^dq}{(2\pi)^d} \,\partial_\kappa R_\kappa(q)\,G^2(\kappa;q)
\,\Gamma^{(4)}_{12ll}(\kappa;p,-p,q,-q),
\end{equation}
where we have introduced the self-energy $\Sigma(\kappa;q)$ and
\beq G^{-1}(\kappa,q)=q^2+R_\kappa(q)+\Sigma(\kappa;q). \eeq
In eq.~(\ref{eq:dGamma2}), and later in this paper, we often denote simply by numbers $1,2,$ etc.,
the $O(N)$ indices $i_1, i_2,$ etc.,  in order to alleviate the notation. A diagrammatic illustration of the right hand side of
eq.~(\ref{eq:dGamma2}) is given in fig.~\ref{fig:dGamma2}.
Similarly, the flow of the  the 4-point function in vanishing field reads:
\begin{eqnarray}
\label{4point}
&&\partial_\kappa\Gamma^{(4)}_{1234}(\kappa;p_1,p_2,p_3,p_4)=
\int \frac{d^dq}{(2\pi)^d}\,\partial_\kappa R_k(q^2)\,G^2(\kappa;q) \nonumber \\
&&\hspace{.5cm}\times\left\lbrace
G(\kappa;q+p_1+p_2)\Gamma^{(4)}_{12ij}(\kappa;p_1,p_2,q,-q-p_1-p_2)
\Gamma^{(4)}_{34ij}(\kappa;p_3,p_4,-q,q-p_3-p_4) \right.\nonumber \\
&&\hspace{.80cm}+\,G(\kappa;q+p_1+p_3)\Gamma^{(4)}_{13ij}(\kappa;p_1,p_3,q,-q-p_1-p_3)
\Gamma^{(4)}_{24ij}(\kappa;p_2,p_4,-q,q-p_2-p_4) \nonumber \\
&&\hspace{.80cm}\left.+\,G(\kappa;q+p_1+p_4)\Gamma^{(4)}_{14ij}(\kappa;p_1,p_4,q,-q-p_1-p_4)
\Gamma^{(4)}_{32ij}(\kappa;p_3,p_2,-q,q-p_3-p_2) \right\rbrace \nonumber \\
&&\hspace{2cm}-\,\frac{1}{2}\int
\frac{d^dq}{(2\pi)^d}\partial_\kappa R_\kappa(q)G^2(\kappa;q)
\Gamma^{(6)}_{1234ii}(\kappa;p_1,p_2,p_3,p_4,q,-q) .
\end{eqnarray}
The four contributions in the r.h.s.~of eq.~(\ref{4point}) are represented in the
diagrams shown in figs.~\ref{fig:dGamma4} and \ref{fig:dGamma4_6}.

Eqs.~(\ref{eq:dGamma2}) and (\ref{4point}) for the 2- and 4-point functions constitute the beginning
of an infinite hierarchy
of exact equations for the $n$-point functions, with the flow equation for the $n$-point function
involving all the $m$-point functions up to $m=n+2$.  Clearly, solving this hierarchy requires approximations. In the rest of this section we discuss various approximations that are commonly used in the context of the NPRG, and that we shall exploit in the more general scheme presented in the next section. In the next sub-section we recall how perturbation theory can be
recovered from the hierarchy  through an iterative procedure. Then, we focus on the regime of small momenta  where an expansion in powers of gradients of the field often yield accurate results. In particular we  briefly discuss the leading order of this expansion, the Local Potential Approximation (LPA). Finally, in the last subsection, we review simple properties of correlations functions of the O($N$) model in the limit of large $N$: this will provide a simple, yet non trivial,  example  in which the momentum dependence of  correlation functions can be analyzed in detail.

\subsection{Perturbation Theory}

\label{pert-theory}

Perturbation theory can be recovered by solving the exact flow
equations iteratively, starting with the classical action as
initial input (see e.g. \cite{Polchinski,Bonini}). The perturbative expansion of the effective action, or equivalently the loop expansion,  is controlled by a ``small  parameter '', namely $\hbar$. Making  this parameter explicit one
rewrites eq.~(\ref{mastereq}) as
\beq\label{mastereqh}
\partial_\kappa \Gamma_\kappa[\phi]=\frac{\hbar}{2}\Tr \partial_\kappa R_\kappa
\left[\Gamma_\kappa^{(2)}+R_\kappa\right]^{-1}, \eeq and then
proceed to the expansion in powers of $\hbar$. In leading
order $\Gamma_\kappa[\phi]$ is independent of $\kappa$ and is equal to the classical action: \beq
\Gamma^{[0]}_\kappa[\phi]=S[\phi]+\mathcal{O}( \hbar) . \eeq The solution of this equation can be used to obtain  an approximation for the 2-point function $\Gamma_\kappa^{(2)}[\phi]$, by taking the second derivative of  $S[\phi]$ with respect to $\phi$ (see eq.~(\ref{rednptfcts})).  At
the next iteration this is used in the right hand side of eq.~(\ref{mastereqh}) in order to obtain the order one correction to $\Gamma_\kappa[\phi]$. One gets then, after integrating the flow
equation from $\Lambda$ to $\kappa$ (using the fact that $S^{(2)}$ is independent of
$\kappa$) \beq \Gamma^{[1]}_\kappa[\phi]&=&S[\phi]+\frac{\hbar}{2}\Tr
\log \left[\frac{S^{(2)}+R_\kappa}{S^{(2)}+R_{\Lambda}}\right]
+\mathcal{O}(\hbar^2), \eeq where one recognizes the familiar one-loop expression of the effective action. One can repeat the procedure and show
that, after $n$ iterations, one reproduces the result that one would obtain by calculating $\Gamma_\kappa[\phi]$ using perturbation theory at  order $n$-loop (with the IR cut-off).

In the case of massless theories,  which we are interested in here, this iteration scheme is  applicable only  for values of $\kappa$ not too small.
Indeed, in general, perturbation theory stops to make sense \cite{Zinn-Justin:2002ru} when $\kappa\simle \kappa_c$, with $\kappa_c\sim u^{1/(4-d)}$ , where $u$ is the coupling constant defined in eq.~ (\ref{classicalaction}) (concrete estimates of $\kappa_c$ will be presented in the next subsection). When $\kappa \to 0$, perturbative calculations may lead to  infrared divergent expressions. This difficulty is particularly important in the scaling regime, where $p \ll p_c \sim \kappa_c$. Other approximation schemes are then required.

\subsection{The local potential approximation}

\label{LPA}

The derivative expansion  offers the possibility to calculate some properties of the scaling regime. It
exploits the fact that the shape of the regulator in the flow
equations (e.g. eqs.~(\ref{eq:dGamma2}) or (\ref{4point})) forces
the loop momentum $q$ to be smaller than $\kappa$, i.e., only
momenta $q\simle \kappa$ contribute to the flow. Besides, in general, the
regulator insures  that, as long as $\kappa\ne 0$,  all vertices are smooth functions of
momenta \footnote{The smoothness of $n$-point functions would be spoiled
by a sharp cut-off regulator. The Litim regulator used in this paper would also lead to difficulties in high orders; however it allows for a Taylor expansion of n-point functions
at leading orders in $q^2/\kappa^2$, which is enough for our
purposes.}. Then, in the calculation of the $n$-point functions at
small external momenta $p_i$, it is  possible to expand the $n$-point functions
in the r.h.s. of the flow equations in terms of $q^2/\kappa^2$ and
$p_i^2/\kappa^2$, or equivalently in terms of the derivatives of
the field. Note, however, that since eventually $\kappa\to 0$, such
an expansion strictly makes sense only for $p_i=0$, unless there is a
mass in the problem.

 In leading order, this procedure reduces to the so-called local potential approximation (LPA), which assumes that the effective action has the form:
\begin{equation}\label{gammaLPA}
\Gamma_\kappa^{LPA}[\phi]=\int d^dx \left\{
\frac{1}{2}\partial_{\mu}\phi_i\partial_{\mu}\phi_i+V_\kappa(\rho)\right\},
\end{equation}
where $\rho\equiv \phi_i\phi_i/2$.
The derivative term here is simply the one appearing in the
classical action, and $V_\kappa(\rho)$ is the effective potential.
The exact flow equation for $V_\kappa$ is
easily obtained  by assuming that the field $\phi$ is constant in
eq.~(\ref{mastereq}). One needs however to take into account the $O(N)$ symmetry,
and to decompose the propagator of the scalar field in  a constant background
$\phi_i$ into its transverse ($G_T$) and
longitudinal ($G_L$)  components:
\beq
G_{ij}(\kappa;q)=G_T(\kappa;q)\left(\delta_{ij}-\frac{\phi_i\phi_j}{2\rho}\right)
+G_L(\kappa;q)\frac{\phi_i\phi_j}{2\rho}.
\eeq
Then the equation for the potential reads:
 \beq\label{eq-pot}
\partial_\kappa V_\kappa(\rho)=\frac{1}{2}\int
\frac{d^dq}{(2\pi)^d} \partial_\kappa R_\kappa(q) \left\lbrace
(N-1)G_T(\kappa;q)+G_L(\kappa;q)\right\rbrace .
\eeq
By using the LPA effective action, eq.~ (\ref{gammaLPA}), one gets
\begin{eqnarray}
G_T(\kappa;q)&=&\frac{1}{q^2+V'(\rho)+R_k(q)} ,\nonumber \\
G_L(\kappa;q)&=&\frac{1}{q^2+V'(\rho)+2\rho V''(\rho)+R_\kappa(q)},
\end{eqnarray}
with $V'(\rho)=dV/d\rho$ and $V''(\rho)=d^2V/d\rho^2$. With these propagators, eq.~(\ref{eq-pot}) becomes then a closed equation.

Higher order corrections to the LPA include terms in the effective
action with an increasing number of derivatives. Although there is
no formal proof of convergence,
 the derivative expansion exhibits quick apparent
convergence if the regulator $R_\kappa(q)$ is appropriately
chosen  \cite{Litim,Canet02,Canet03}. In practice, the LPA
reproduces well the physical quantities dominated by  small
momenta (such as the effective potential or critical exponents) in
all theories where it has been tested (see, for example,
\cite{Berges02,Canet04}). Higher order corrections
lead to significant improvements \cite{Bervi}, and the derivative expansion has been pushed up to third
order  \cite{Canet03}, yielding  critical exponents in the Ising
universality class of the same level of precision as those obtained with the best
accepted methods (see e.g. \cite{Canet03}).

An interesting  improvement of the LPA, which we refer to as the
LPA',    takes into account   a running field renormalisation
constant $Z_\kappa$ and allows for a non trivial anomalous
dimension, determined from the cut-off dependence of $Z_\kappa$
\cite{Wetterich93}. In the LPA', the effective action is assumed to be of the form:
\beq
\Gamma^{LPA'}_\kappa[\phi] =\int
d^dx\left\lbrace
\frac{Z_\kappa}{2}\partial_{\mu}\phi_i\partial_{\mu}\phi_i+V_\kappa(\rho)\right\rbrace.
\label{LPA'}
\eeq
where  $Z_\kappa$ is a function of $\kappa$ (and
not of $\rho$). It is useful to explicitly include the field normalisation in the regulator (\ref{reg-litim}), i.e., we redefine $R_\kappa$ by multiplying it by the
factor $Z_\kappa$. Thus, the regulator used in the present work becomes:
\beq \label{defRk}
R_\kappa(q)=Z_\kappa\, (\kappa^2-q^2)\Theta(\kappa^2-q^2).
\eeq
The factor $Z_\kappa$ is determined from the flow equation for $\Gamma_\kappa^{(2)}$
in a constant external field, which can be derived from
eq.~(\ref{mastereq}). The vertices and propagators entering  this equation are those
dictated by
the form (\ref{LPA'}) assumed for the effective action. By expanding the resulting equation to order $p^2$:
$\Gamma^{(2)}(\kappa;p) - \Gamma^{(2)}(\kappa;0) \sim p^2 Z_\kappa$ (recall that for non vanishing $\kappa$, $\Gamma^{(2)}(\kappa;p)$ is a smooth function of $p$), one obtains the following
equation for $Z_\kappa$ \cite{Berges02}:
\begin{eqnarray}
\label{equationpoureta}
\partial_\kappa Z_\kappa=\frac{4}{d}\rho_0 (V''(\rho_0))^2
\tilde{\partial}_\kappa\int \frac{d^dq}{(2\pi)^d} q^2G_L^2(\kappa;q)
G_T^2(\kappa; q) (Z_\kappa+R_\kappa'(q))^2,
\end{eqnarray}
where $R_\kappa'(q)\equiv \del R_\kappa(q)/\del q^2$, the derivative $\tilde{\partial}_\kappa$ acts only on the explicit  factors  $R_\kappa$
(and their derivatives),  and  $\rho$ is fixed at its
running minimum $\rho=\rho_0$ (which depends on $\kappa$).
The anomalous dimension is  related to $Z_\kappa$ by (see e.g.  \cite{Berges02}; for a simple proof, see app.~\ref{appendixeta}):
\beq\label{defZk}
\eta_\kappa=-\kappa\partial_\kappa \ln Z_\kappa .
\eeq
In the LPA'  the flow equation for the effective potential is the same as in the LPA, eq.~(\ref{eq-pot}), except for the replacement $q^2\rightarrow Z_\kappa q^2$ in the propagators. It follows that the flow equation for the potential is coupled with the flow equation for $Z_\kappa$, eq.~(\ref{equationpoureta}).

The  derivatives of $V_\kappa(\rho)$ with respect to $\rho$ give
the $n$-point functions at {\it zero external momenta}  as a function of
$\kappa$. We shall be mostly concerned in this paper with the
critical regime where $\rho_0(\kappa=0)=0$, and hence in $n$-point
functions in vanishing external field, for which we shall
introduce special notation. We set: \beq\label{defcst} m_\kappa^2
\equiv \left.\frac{dV_\kappa}{d\rho}\right|_{\rho=0} , \hskip 1 cm
g_\kappa\equiv\left.\frac{d^2V_\kappa}{d\rho^2}\right|_{\rho=0} ,
\hskip 1 cm h_\kappa
\equiv \left.\frac{d^3V_\kappa}{d\rho^3}\right|_{\rho=0} . \eeq For
vanishing external field the propagator is diagonal,
$G_{12}(\kappa;q) = \delta_{12} G_{LPA'}(\kappa;q)$, with
\beq\label{propLPA'} G_{LPA'}^{-1}(\kappa;q)=Z_\kappa
q^2+R_\kappa(q)+m_\kappa^2. \eeq For the $n$-point functions
$\Gamma^{(4)}$ and $\Gamma^{(6)}$, we have, respectively:
\beq\label{gamma4LPA} \Gamma_{1234}^{(4)\,
LPA'}(\kappa)=g_\kappa\,(\delta_{12}\;\delta_{34} +
\delta_{13}\;\delta_{24}+\delta_{14}\;\delta_{23} ) \; , \eeq and
\begin{eqnarray}\label{gamma6LPA}
\Gamma^{(6)\,LPA'}_{123456}(\kappa)&=&
h_\kappa\,\left[\delta_{56}\left(\delta_{12}\delta_{34}+\delta_{13}\delta_{24}
+\delta_{14}\delta_{23}\right)+\delta_{46}\left(\delta_{12}\delta_{35}+\delta_{13}\delta_{25}
+\delta_{23}\delta_{15}\right) \right.\nonumber\\
&&\hspace{+0.5cm}
+\delta_{36}\left(\delta_{12}\delta_{45}+\delta_{14}\delta_{25}
+\delta_{15}\delta_{24}\right)+\delta_{26}\left(\delta_{13}\delta_{45}+\delta_{14}\delta_{35}
+\delta_{15}\delta_{34}\right) \nonumber \\
&&\hspace{.5cm}\left.+\delta_{16}\left(\delta_{23}\delta_{45}+\delta_{24}\delta_{35}
+\delta_{25}\delta_{34}\right)\right]\, .
\end{eqnarray}

In order to factor out the  large variations of the effective potential which arise when
$\kappa$ varies from the microscopic scale $\Lambda$ to the
physical scale $\kappa=0$, and also to exhibit the fixed point structure,  it is convenient to isolate  the
explicit scale factors ($V_\kappa\sim \kappa^d$, $Z_\kappa\rho\sim \kappa^{d-2}$) and to   define dimensionless quantities:
\beq
v_\kappa (z)\equiv K_d^{-1} \kappa^{-d} V_\kappa
(\rho),
\eeq
with
\beq
z \equiv K_d^{-1} \; Z_\kappa \; \kappa^{2-d} \;
\rho.
\eeq
In these definitions, for further simplifications,  we have also included a factor
$K_d$, which originates from angular integrations:
\beq
K_d^{-1}\equiv\;
2^{d-1}\; \pi^{d/2} \; d\;\Gamma(d/2).
\eeq
Note that $K_d$ can be a small number,
e.g. $K_3=1/6\pi^2$. We also introduce dimensionless couplings:
\beq\label{adimcons}
m_\kappa^2\equiv Z_\kappa \kappa^2\, \hat
m_\kappa^2, \hskip 1 cm g_\kappa \equiv K_d^{-1} Z_\kappa^2
\kappa^{4-d}\, \hat g_\kappa, \hskip 1 cm h_\kappa \equiv  K_d^{-2}
Z_\kappa^3 \kappa^{6-2d} \,\hat h_\kappa,
\eeq
so that:
\beq
\hat m_\kappa^2= \left.\frac{dv_\kappa}{dz}\right|_{z=0}\hskip 1
cm \hat g_\kappa=\left.\frac{d^2v_\kappa}{dz^2}\right|_{z=0}
\hskip 1 cm \hat
h_\kappa =\left.\frac{d^3v_\kappa}{dz^3}\right|_{z=0}.
\eeq

The  solution of the LPA' is well
documented in the literature (see e.g. \cite{Berges02,Canet02}).
It is convenient to
solve the equation for the derivative of the potential
 with respect to $z$, i.e., $w_\kappa(z)\equiv \partial_z
v_\kappa(z)$, rather than that for the effective potential itself. With the Litim regulator (\ref{defRk}), the integrals in eqs.~(\ref{eq-pot})
and (\ref{equationpoureta}) can be done analytically. One gets:
\beq\label{flow_of_w}
\kappa\partial_\kappa w_\kappa\!=\! -(2\!-\!\eta_\kappa) w_\kappa + (d\!-\!2\!+\!\eta_\kappa) z w'_\kappa
 -\! \left( 1\!-\!\frac{\eta_\kappa}{d+2} \right) \!\left( \frac{(N\!-\!1) w'_\kappa}{(1+w_\kappa)^2} \!+\!
 \frac{3w'_\kappa + 2 zw''_\kappa}{(1+w_\kappa+2 z w'_\kappa)^2} \right),\nonumber\\
\eeq
and
\beq\label{eq.eta}
\eta_\kappa = \frac{4 z_0 (w'_\kappa(z_0))^2}{(1+2z_0 w'_\kappa(z_0))^2} \; ,
\eeq
where $w'_\kappa=\partial_z w_\kappa(z)$, $w''_\kappa=\partial_z^2 w_\kappa(z)$, and
 $z_0=z_0(\kappa)$ is the running minimum of the potential ($w_\kappa(z_0)=0$). Eqs.~(\ref{flow_of_w}) and (\ref{eq.eta}) are solved starting from the initial condition at
$\kappa=\Lambda$:
\beq\label{wLambdadez}
w_\kappa(z,\kappa=\Lambda) = \hat m_\Lambda^2 + \hat g_\Lambda z,
\eeq
where $ \hat m_\Lambda$ and $ \hat g_\Lambda$ are related to the parameters $r$ and $u$ of the classical action (\ref{classicalaction}) by
\beq\label{relclas}
 \hat m_\Lambda^2 =\frac{r}{\Lambda^2}\qquad \hat g_\Lambda = \frac{u }{\Lambda^{4-d}}\frac{K_d}{3}.
\eeq

Before looking at some results obtained by solving numerically
eqs.~(\ref{flow_of_w}) and (\ref{eq.eta}), it is useful to get
insight into the expected behaviour of the solution by solving
eq.~(\ref{flow_of_w}) approximately  \cite{Berges02}, ignoring the anomalous
dimension. To this aim, we assume that, for all $\kappa$,
$w_\kappa(z)$ retains  the form of eq.~(\ref{wLambdadez}), i.e.,
 \beq\label{formwap}
w_\kappa(z)=\hat m_\kappa^2+ \hat g_\kappa\, z .
\eeq
The minimum of the potential, $z_0(\kappa)$,
satisfies $w_\kappa(z_0)=0$, i.e.,  $z_0(\kappa)=-\hat m_\kappa^2/ \hat g_\kappa$. The equations
for $z_0(\kappa)$ and $\hat g_\kappa$ are easily obtained from eq.~(\ref{flow_of_w}),
taking into account that,  at criticality, $z_0 \hat g\ll 1$ to make simplifications whenever appropriate. One gets:
\begin{eqnarray}\label{LPAapprox}
\kappa\frac{d z_0}{d\kappa}&=&-(d-2)\,z_0+ N+2-6z_0\hat g_\kappa,
 \nonumber \\
\kappa\frac{d \hat g_\kappa}{d\kappa}&=&(d-4)\,\hat g_\kappa+  2(N+8) \,\hat g_\kappa^2. \
\end{eqnarray}

The  equation for $\hat g_\kappa$ defines the usual one-loop
$\beta$-function; in this approximation this equation decouples and can be solved explicitly:
\beq\label{eq:hatg} \hat g_\kappa=\frac{\hat
g^*}{1+\left(\frac{\kappa}{\kappa_c}\right)^{4-d}} , \eeq where
$\hat g^*$ is the value of $\hat g$ at the IR fixed point,  $\hat
g^*=(4-d)/{(2(N+8))}$,  and $\kappa_c$ the value of $\kappa$ for
which $\hat g_\kappa=\hat g^*/2$. We have ($\hat g^*\gg \hat
g_\Lambda$): \beq\label{kappac}
\left(\frac{\kappa_c}{\Lambda}\right)^{d-4}=\frac{\hat g^*-\hat
g_\Lambda}{\hat g_\Lambda} \approx\frac{\hat g^*}{\hat g_\Lambda}
. \eeq $\kappa_c^{4-d}=uK_d/(3 \hat g^*)$ is the typical  scale which separates the
scaling region, dominated by the IR fixed point, where $\hat g= \hat g^*$, from the
perturbative region,  dominated by the UV fixed point $\hat g=0$
(when $\kappa\gg \kappa_c$, one can expand $\hat g_\kappa$ in
powers of $\kappa_c/\kappa$; in leading order
$g_\kappa=(u/3)(1-(\kappa_c/\kappa)^{4-d})$).

\begin{figure}[t!]
\begin{center}
\includegraphics*[width=10cm]{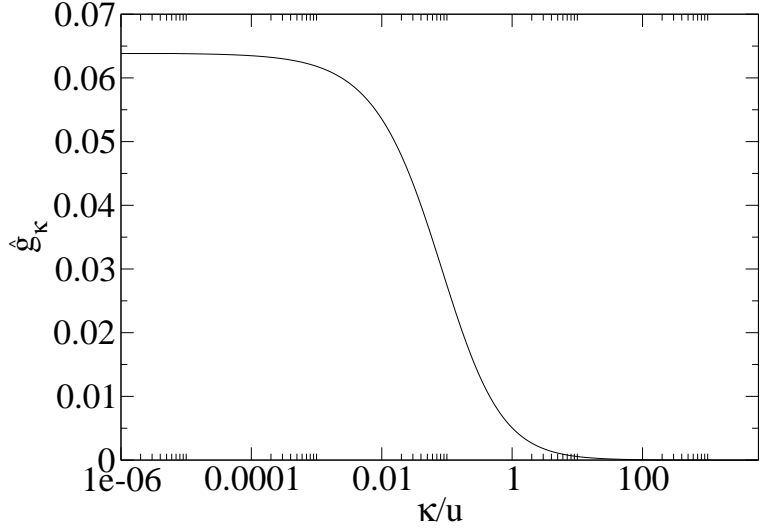}
\end{center}
\caption{\label{lambda} The dimensionless coupling $\hat g_\kappa$ as a function of
$\kappa/u$ (in a logarithmic scale) obtained by solving the LPA' equations for $N=2$ and $d=3$. The value of
$\hat g_\kappa$ at the IR fixed point is $\hat g^*=0.064$. The value of $\kappa$ for which $\hat g_\kappa= \hat g^*/2$ is $\kappa_c=0.072$.}
\end{figure}

\begin{figure}[t!]
\begin{center}
\includegraphics*[width=10cm]{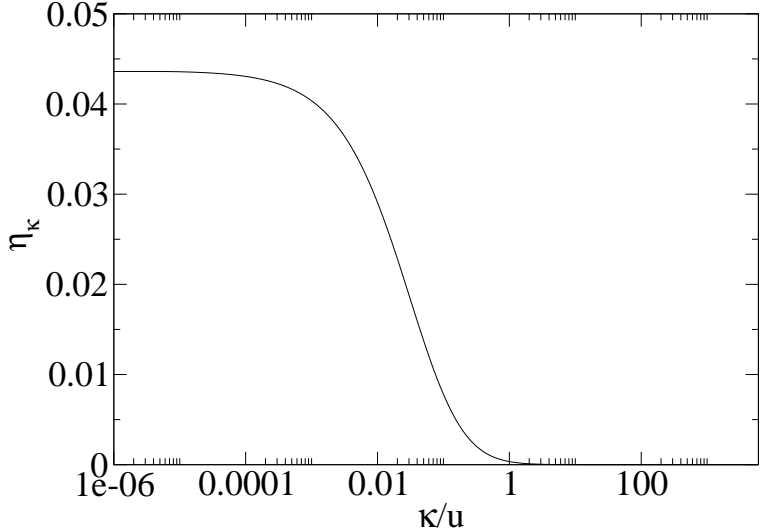}
\end{center}
\caption{\label{eta} The anomalous dimension $\eta_\kappa$ as a function of
$\kappa/u$ (in a logarithmic scale) obtained by solving the LPA' equations for $N=2$ and $d=3$. The value of
$\eta$ at the IR fixed point is $\eta^*=0.044$.}
\end{figure}

We show in figs.~\ref{lambda} and \ref{eta} the dimensionless
coupling $\hat g_\kappa$ and the anomalous dimension $\eta_\kappa$
obtained by solving the complete LPA' equations numerically for
$d=3$ and $N=2$. The coupling constant $ \hat g_\Lambda$ has been
fixed to a small value, and   $\hat m_\Lambda^2$  has been
adjusted in order to reach the IR fixed point as $\kappa\to 0$.
Note that $w_\kappa$ depends a priori on $u$, $\kappa$, and
$\Lambda$, but since it is dimensionless, it can  only  depend  on
the ratios $\kappa/u$ and $u/\Lambda$ (in $d=3$).
 However, because the theory is super-renormalisable in $d=3$, $w_\kappa$ becomes  independent of
$u/\Lambda$ in the limit of large $\Lambda$. One finds numerically that this regime is
attained when $u/\Lambda \simle 10^{-3}$,
 a condition satisfied in all numerical results presented in this paper:  more precisely,
we used $u/\Lambda=3\; K_3^{-1}\;
10^{-6} \simeq 1.8\; 10^{-4} $, i.e., $\hat g_{\kappa=\Lambda} = 10^{-6}$; the corresponding value of $\hat m_\Lambda^2$ needed to reach the fixed point is $\hat m_\Lambda^2=-3.999527\cdots\times 10^{-6}$.
The general behaviors seen in figs.~\ref{lambda} and \ref{eta} are those expected from the approximate  analytic solution discussed above, in particular the
fixed point values reached at small
$\kappa$. On a logarithmic scale, the change of regime between the perturbative regime at large
$\kappa$, and the scaling regime at
small $\kappa$ occurs
rather rapidly, at the typical scale $\kappa_c$. In fig.~\ref{lambda}
$\kappa_c/u \sim 0.07$, not far from  the value obtained in the approximate analysis presented above:
from eqs.~(\ref{relclas})  and (\ref{kappac}), for $d=3$ and $N=2$, $\kappa_c/u =20 K_3/3 \simeq 0.11$.

Before closing this subsection, let us write the flow equations
for the 2- and 4-point functions in vanishing external field, in
the LPA' limit, in a form that we shall  use later. These
equations are obtained by deriving once and twice
eq.~(\ref{eq-pot}) with respect to $\rho$, then setting $\rho=0$,
and using the definitions in eq.~(\ref{defcst}). They read, respectively: \beq\label{preMk}
\kappa\partial_\kappa m_\kappa^2 = -\frac{(N+2)}{2} g_\kappa I_d^{(2)} ,
\eeq
and
\beq\label{preFk}
\kappa\partial_\kappa g_\kappa =
(N+8) g_\kappa^2 I_d^{(3)}(\kappa) - \frac{1}{2}
(N+4)\; h_\kappa I_d^{(2)}(\kappa) ,
\eeq
where we have defined
\beq\label{Ink}
I_d^{(n)}(\kappa) &\equiv &\int
\frac{d^dq}{(2\pi)^d}\kappa\partial_\kappa R_\kappa(q^2)G_{LPA'}^n(\kappa; q)\nonumber\\
 &= &2 K_d \frac{\kappa^{d+2-2n}}{Z_\kappa^{n-1}}\frac{1}{(1+\hat m^2_\kappa)^n}
\left(1-\frac{\eta_\kappa}{d+2}\right) , \eeq the explicit
form in the second line being obtained for the Litim regulator. Note that, after going to dimensionless variables and making the same approximations that leads
us to eqs.~(\ref{LPAapprox}) (neglect the second derivative of $w_\kappa$ with respect to
$z$, and assume  $|\hat m^2_\kappa| \ll 1$) one can transform eq.~(\ref{preFk}) into the second of
eqs.~(\ref{LPAapprox}).

\begin{figure}[t!]
\begin{center}
\includegraphics*[width=10cm]{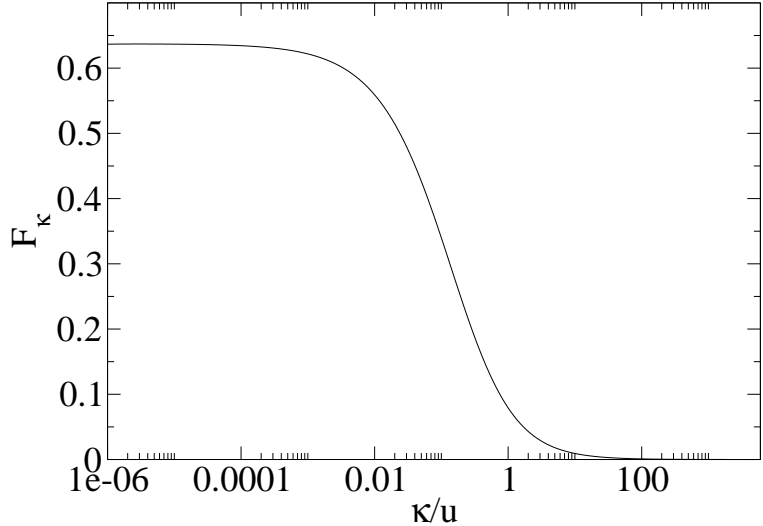}
\end{center}
\caption{\label{Fk} The function $F_\kappa$ in eq.~(\ref{gderivtotale}) as a function of $\kappa/u$ (in a logarithmic scale), calculated for $N=2$ and $d=3$.}
\end{figure}

For further use, we also rewrite eq.~(\ref{preFk}) in the following  form:
\beq\label{gderivtotale}
\kappa\del_\kappa g_\kappa = (N+8)g_\kappa^2
I_d^{(3)}(\kappa) (1-F_\kappa) .
\eeq
where
\beq\label{defFk}
F_\kappa=\frac{1}{2} \frac{N+4}{N+8}\frac{I_d^{(2)}(\kappa)}{I_d^{(3)}(\kappa)}\frac{h_\kappa}{g_\kappa^2}=
\frac{1}{2}\frac{N+4}{N+8} (1+\hat m_\kappa^2) \frac{\hat h_\kappa}{\hat
g_\kappa^2}
\eeq
The function $F_\kappa$ gives a measure of the relative magnitude of the contribution of the 6-point vertex  term  in the flow equation for the 4-point function.
 One can see on fig.~\ref{Fk} that,  as expected, the relative contribution of the 6-point vertex
is negligeable in the perturbative regime ($\kappa\gg \kappa_c$), but becomes of order 1 in the scaling regime ($\kappa\ll\kappa_c$).

\subsection{Correlation functions at large N}

\label{largeN}

In the critical case, the derivative expansion gives accurate results for the
correlation functions and their derivatives only at zero external
momenta. In order to get insight into the effect of non vanishing
external momenta we consider now  the correlation functions in the
large $N$ limit (at fixed $uN$). Our goal here is to illustrate some general features of the momentum dependence of the  correlation functions,  and how this is affected by the regulator,  not to present a consistent discussion of the flow equations and their solutions, which can be found in the literature  \cite{Ellwanger94,D'Attanasio97,Tetradis95,PLB}.  Thus we shall not attempt to solve directly the NPRG equations: since they do not close, their solution requires a somewhat elaborate treatment (see e.g. \cite{PLB}). Rather, we shall simply write the solution for the first $n$-point functions, relying on well known results \cite{Moshe:2003xn}, and verify that they do satisfy the NPRG equations.

For vanishing field, the inverse propagator is of the form
\beq\label{proplargeN}
 G^{-1}(\kappa; q)=q^2+m_\kappa^2+R_\kappa(q),
\eeq
where $m_\kappa$ is a running mass given by a gap equation
\beq
\label{gap}
m_\kappa^2=r+\frac{Nu}{6}\int \frac{d^dq}{(2\pi)^d}\,
(G(\kappa;q)-G(\Lambda;q)).
\eeq
The 4-point function has the following structure:
\beq\label{4pointlargeN}
\Gamma_{1234}^{(4)}(\kappa;p_1,p_2,p_3,p_4)=\delta_{12}\delta_{34}g_\kappa(p_1+p_2)
+\delta_{13}\delta_{24}g_\kappa(p_1+p_3)+\delta_{14}\delta_{23}g_\kappa(p_1+p_4),
\eeq
where
 $g_\kappa(p)$ is given by
\begin{equation}\label{B3}
g_\kappa(p)=\frac{u}{3} \frac{1}{1+\frac{Nu}{6}B_d(\kappa;p)},
\end{equation}
with
\beq
B_d(\kappa;p)\equiv \int\frac{d^dq}{(2\pi)^d}\,
G(\kappa;q)G(\kappa;p+q).
\eeq
Finally we shall need shortly the 6-point function $\Gamma_{1234mm}^{(6)}(\kappa; p_1,p_2,p_3,p_4,q,-q)$ (summation over repeated indices is understood)
\begin{eqnarray}\label{gamma6N}
&&\hspace{-.5cm} \frac{1}{N}\Gamma_{1234mm}^{(6)}(\kappa; p_1,p_2,p_3,p_4,q,-q)\nonumber\\
&&=  h_\kappa(p_1+p_2)\delta_{12}\delta_{34}
+h_\kappa(p_1+p_3)\delta_{13}\delta_{24}+h_\kappa(p_1+p_4)\delta_{14}\delta_{23},
\end{eqnarray}
with
\beq\label{Hkappa}
h_\kappa(p)=N g_\kappa(0)g^2_\kappa(p)\int\frac{d^dq}{(2\pi)^d}G^2(\kappa;q)G(\kappa;q+p).
\eeq

All these results can be obtained in a straightforward fashion by calculating the corresponding Feynman diagrams with a regulator. It is however easy to  verify that the various $n$-point functions that we have just written are indeed   solutions of the flow equations in the large $N$ limit.

 To this aim, one notes first that eq.~(\ref{eq:dGamma2}) reduces to an equation for the running mass:
\beq \label{deriveedemk2}
\partial_\kappa m_\kappa^2=-\frac{1}{2}Ng_\kappa(0)\int\frac{d^dq}{(2\pi)^d}\partial_\kappa R_\kappa(q)G^2(\kappa;q) ,
\eeq
and using eq.~(\ref{B3}), it is easy  to check that this equation is equivalent to the gap equation,  eq.~(\ref{gap}).

Next, we observe that in the large $N$ limit, a single channel contributes in eq.~(\ref{4point}) for the $4$-point function;  one can then
use the following identity in this limit:
\begin{eqnarray}\label{B7}
\Gamma^{(4)}_{12ij}(\kappa;p_1,p_2,q,\!-\!q\!-\!p_1\!-\!p_2)\Gamma^{(4)}_{34ij}(\kappa;p_3,p_4,\!-\!q,q\!-\!p_3\!-
\!p_4)
=Ng^2_\kappa(p_1+p_2)\delta_{12}\delta_{34},
\end{eqnarray}
together with eq.~(\ref{gamma6N}) for $\Gamma^{(6)}$, and obtains:
\beq\label{B8}
\kappa\partial_\kappa g_\kappa(p)=Ng_\kappa^2(p) J^{(3)}_d(\kappa;p)-\frac{N}{2}h_\kappa(p)I^{(2)}_d(\kappa),
\eeq
where the function $I^{(2)}_d(\kappa)$ is that defined  in
eq.~(\ref{Ink}), with here  $n=2$  and the
propagator (\ref{proplargeN}) replacing $G_{LPA'}$. The function $J_d^{(3)}(\kappa;p)$
is obtained from the general definition
\beq
\label{Jnk} J_d^{(n)}(\kappa;p)\equiv\int \frac{d^d q}{(2\pi)^d}
\kappa\partial_\kappa
R_\kappa(q)G^{n-1}(\kappa;q)G(\kappa;p+q).
\eeq
 Note that
$J_d^{(n)}(\kappa;p=0)=I_d^{(n)}(\kappa)$. Explicit expressions for the function $J_3^{(3)}(\kappa;p)$ are given in  app.~\ref{functionJ}.

 At this point we remark  that the flow equation for $g_\kappa(p)$
can also be obtained directly from the explicit expression
(\ref{B3}), in the form:
\beq \label{deriveedegk}
\partial_\kappa g_\kappa(p)=-\frac{N}{2}g^2_\kappa(p)\partial_\kappa\int\frac{d^dq}{(2\pi)^d}G(\kappa;q)G(\kappa;q+p).
\eeq
 It is then
straightforward to verify, using eqs.~(\ref{deriveedemk2}) and
(\ref{Hkappa}) that eqs.~(\ref{B8}) and (\ref{deriveedegk}) are
indeed equivalent. The first term
in eq.~(\ref{B8}) comes from the derivative of the cut-off
function in the propagators in eq.~(\ref{deriveedegk}), while the
second term, which involves the 6-point vertex, comes from  the
derivative of the running mass in the propagators.

Note that eqs.~(\ref{deriveedemk2})  for $m_\kappa$ and (\ref{B8}) for $g_\kappa(p=0)$ become identical respectively to eqs.~(\ref{preMk}) and (\ref{preFk}) of the LPA in the large $N$ limit, a well know property \cite{D'Attanasio97}.

In view of the approximations that
we shall develop in the next section, it is worth  analyzing
characteristic features of the function $g_\kappa(p)$. For simplicity we specialize for the  rest of this subsection to $d=3$. Furthermore, for the purpose of the present, qualitative, discussion, one may assume
$m_\kappa=0$. This allows us to obtain easily
$g_\kappa(p)$ from eq.~(\ref{B3}) in the two limiting cases $\kappa=0$ and $p=0$. In
the first case, we have
\beq\label{B3c}
g_\kappa(0)=\frac{u}{3}\frac{1}{1+\frac{uN}{9\pi^2}\frac{1}{\kappa}}.
\eeq
This is identical to eq.~(\ref{eq:hatg}), with here $\hat
g^*= 1/(2N)$ and $\kappa_c=Nu/9\pi^2$ . (The corresponding
expressions for eq.~(\ref{eq:hatg}) involve $N+8$ instead of $N$,
so that the values of $\hat g^*$ and $\kappa_c$ obtained in the
large $N$ limit may be numerically quite different from the actual
LPA values when $N$ is not too large, e.g. when $N=2$). In the
other case, we have
\begin{equation}\label{B3b}
g_{\kappa=0}(p)=\frac{u}{3}\frac{1}{1+\frac{uN}{48}\frac{1}{p}}=
\frac{u}{3}\frac{p}{p+p_c},
\end{equation}
with $p_c\equiv {uN}/{48}$.

\begin{figure}[t!]
\begin{center}
\includegraphics*[width=10cm,angle=-90]{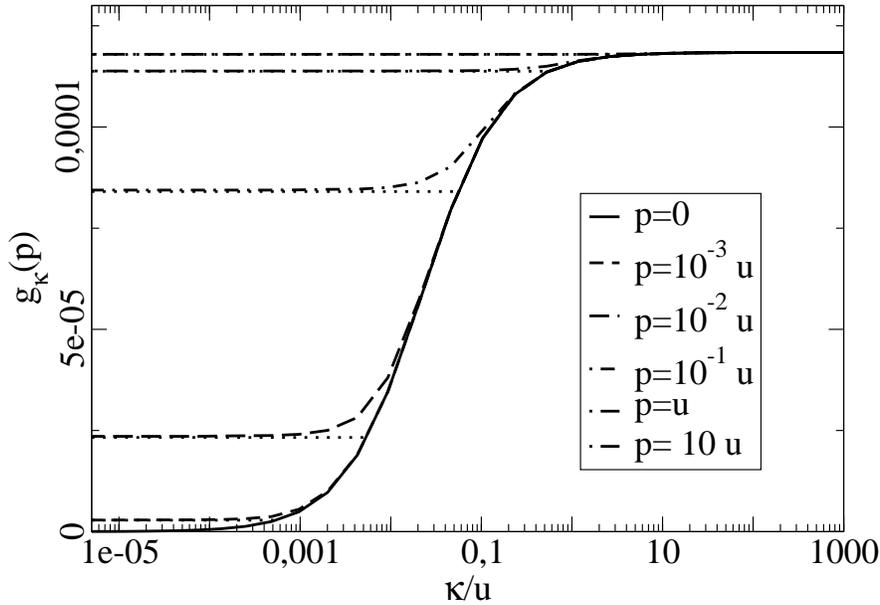}
\end{center}
\caption{\label{fin} The function $g_\kappa(p)$ (in units of $\Lambda$) obtained from a
complete numerical solution of eqs.~(\ref{deriveedemk2}) and
(\ref{B8}), as a function of $\kappa/u$ (in a logarithmic
scale) for five values of $p$: from bottom to top, $p/u=0.001,0.01,0.1,1$ and $10$. The envelope
corresponds to $p=0$. This figure illustrates the decoupling of
modes: for each value of $p$, the flow stops when $\kappa\simle
\alpha p$. The various horizontal asymptotes (dotted llines) correspond to the
single value $\alpha=0.54$. }
\end{figure}

One sees on eqs.~(\ref{B3c}) and (\ref{B3b}) that the dependence
on $p$ of $g_{\kappa=0}(p)$ is quite similar to the dependence on
$\kappa$ of $g_\kappa(p=0)$. In particular both quantities vanish
linearly as $\kappa\to 0$ or $p\to 0$, respectively. The result of
the complete (numerical) calculation, including the effect of the running  mass ( i.e., solving the gap equation (\ref{gap}) and calculating $g_\kappa(p)$ from eq.~(\ref{B3})),  can in
fact be quite well represented (to within a few percents) for arbitrary $p$ and $\kappa$ by the
following approximate formula \beq\label{gapprox}
g_\kappa(p)\approx \frac{u}{3} \frac{X}{1+X}\qquad X\equiv
\frac{\kappa}{\kappa_c}+\frac{p}{p_c}.
\eeq
This simple expression
shows that $p$, when it is non vanishing, plays the same role as
$\kappa$ as an infrared regulator. In particular, at fixed $p$,
the flow  of $g_\kappa(p)$ stops when $X$ becomes independent of $\kappa$, i.e., when $\kappa\simle p(\kappa_c/p_c)$, with $\kappa_c/p_c=16/3\pi^2\approx 0.54$. This
important property of decoupling of the short wavelength modes is
illustrated in fig.~\ref{fin}. As shown by this figure, and also
by the expression (\ref{gapprox}), the momentum dependence of the
4-point function can be obtained from its cut-off dependence at zero momentum. In fact fig.~\ref{fin}
suggests that,  to a very good approximation, there exists a
parameter $\alpha$ such that $g(\kappa;p)\approx g(\kappa;0)$ when
$\kappa>\alpha p$, and $g(\kappa;p)\approx g(\kappa=\alpha p;0)$ when
$\kappa<\alpha p$. From the discussion above, one expects $\alpha\approx \kappa_c/p_c=16/3\pi^2\approx 0.54$, which is indeed in agreement with the analysis in fig.~\ref{fin}.

\begin{figure}[h!]
\includegraphics[width=10cm,angle=-90]{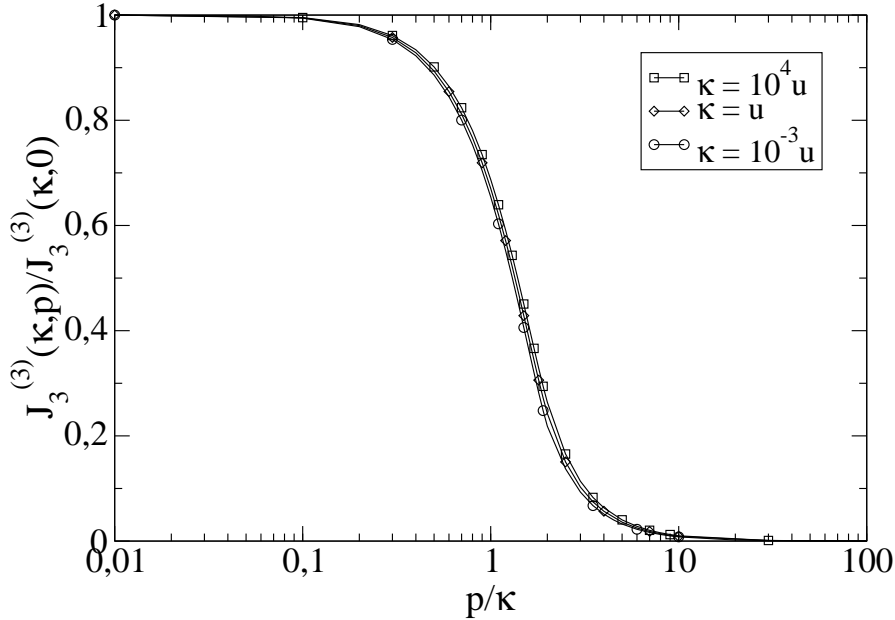}
\caption{\label{fig:theta}
The function $J_3^{(3)}(\kappa;p)/I_3^{(3)}(\kappa)$ as a function of
$p/\kappa$ (in a logarithmic scale), for different values of $\kappa$:
$\kappa=10^{-3} u$ (circles), $\kappa=u$ (diamonds) and
$\kappa=10^3 u$ (squares).}
\end{figure}
\begin{figure}[t!]
\begin{center}
\includegraphics*[width=10cm,angle=-90]{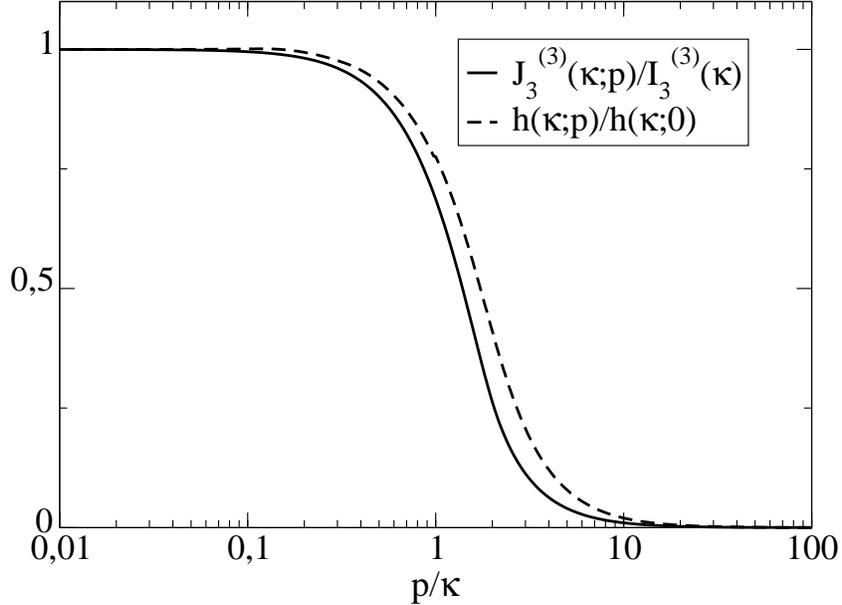}
\end{center}
\caption{\label{H_vs_J} The function $J_3^{(3)}(\kappa;p)/I_3^{(3)}(\kappa)$ as a function of $p/\kappa$ (in a logarithmic scale) (full line). The function $h(\kappa;p)/h(\kappa;0)$ as a function of $p/\kappa$ (dashed line).  }
\end{figure}

\begin{figure}[t!]
\begin{center}
\includegraphics*[width=10cm,angle=-90]{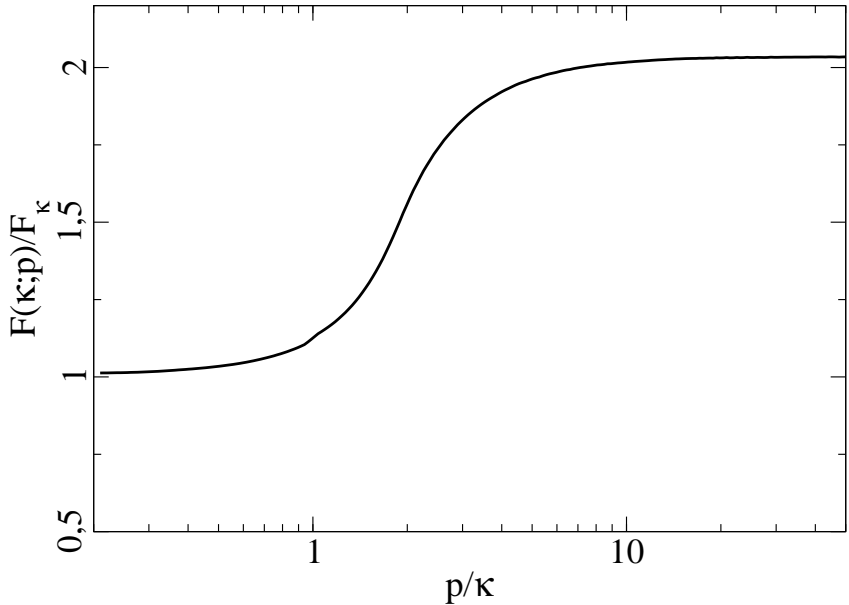}
\end{center}
\caption{\label{Fkp} The function $F(\kappa;p)/F_\kappa$ as a
function of $p/\kappa$ (in a logarithmic scale). }
\end{figure}

In order to understand better the origin of this result,
 we rewrite eq.~(\ref{B8}) as
follows:
\beq\label{B8a}
\partial_\kappa g_\kappa(p)=Ng_\kappa^2(p) J^{(3)}_d(\kappa;p)(1-F(\kappa,p))
\eeq
where \beq\label{gderivtotaleN} F(\kappa;p)\equiv
\frac{1}{2}\frac{h_\kappa(p)I^{(2)}_d(\kappa)}{g_\kappa^2(p)
J^{(3)}_d(\kappa;p)}. \eeq When $p=0$, eq.~(\ref{B8a})
coincides with the LPA equation (\ref{gderivtotale}), and  the
function $F(\kappa;p)$ with the large $N$ limit of the function $F_\kappa$ defined for
the LPA in eq.~(\ref{defFk}). The
$p$-dependence of $J_d^{(n)}(\kappa;p)$ is relatively simple: when
$p\ll \kappa$, $J_d^{(n)}(\kappa;p)\simeq
I_d^{(n)}(\kappa)$; when $p\gg \kappa$, $J_d^{(n)}(\kappa;p)$
vanishes as $1/p^2$. On a logarithmic scale the transition between
these two regimes occurs rapidly at momentum $p\sim \kappa$, as
illustrated on fig.~\ref{fig:theta}. Fig.~\ref{H_vs_J} shows the
similar behavior of the function $h_\kappa(p)/h_\kappa$, where
$h_\kappa(p)$ is the function (\ref{Hkappa}) which appears in the numerator of
eq.~(\ref{gderivtotaleN}). Finally, fig.~\ref{Fkp} displays the
function $F(\kappa;p)/F_\kappa$: as one can see, the momentum
dependence of $F(\kappa;p)$ is non negligible only in the region
where the function $J_d^{(n)}(\kappa;p)$ is negligible, namely for
$\kappa\simge p$. All this suggests that one can rewrite
eq.~(\ref{B8a}) for $g_\kappa(p)$ as follows: \beq\label{B8b}
\partial_\kappa g_\kappa(p)\approx Ng_\kappa^2(p) \Theta(1-\frac{\alpha^2 p^2}{\kappa^2})I^{(3)}_d(\kappa)(1-F_\kappa),
\eeq where $\alpha$ is a parameter of order unity. Eq.~(\ref{B8b})
is just eq.~(\ref{gderivtotale}) in the large $N$ limit, and for
$\kappa>\alpha p$. The $\Theta$-function ensures that the flow
exists only when $\kappa>\alpha p$, and stops for smaller values
of $\kappa$. These are  precisely the features observed in
fig.~\ref{fin}.

\section{Towards the solution of the NPRG equations for arbitrary momenta \label{Approximation_scheme}}

Our proposal to solve the NPRG equations for  the $n$-point functions at arbitrary momenta,  builds upon the lessons
learnt in the specific examples discussed in the previous section.
Namely, we shall take advantage of the decoupling of modes,   exploit
 the solution of the LPA', and use the possibility to increase accuracy through
iterations.

The decoupling of modes is well illustrated in fig.~\ref{fin} of the previous section. It suggests that the momentum dependence of the $n$-point function can be deduced from their $\kappa$-dependence, as obtained from the LPA'. To be more specific, assume for simplicity that all external momenta are of the same order of magnitude, and call them generically $p$. Then as long as $\kappa \simge p$, one can use the LPA' to calculate the $n$-point functions. When $\kappa\simle p$, the flow stops and the $n$-point functions remain at their values for $\kappa\sim p$. Note that this argument ceases to apply when the momenta enter as exceptional configurations, for which effectively $p=0$. These exceptional configurations cause special difficulties  that we shall have to deal with.

 The possibility to increase the accuracy through iterations is based on the property recalled in the previous section, that the iteration of the
NPRG equations, starting with the classical values of the
$n$-point functions as initial input, reconstructs the usual loop
expansion. Thus, one may expect to improve the accuracy of the $n$-point functions at high momenta by iterations.  The situation at small momenta is more subtle. Indeed, in the critical regime, iterations may affect the fixed point structure, and may result in unphysical behaviors. This particular feature will be discussed in \cite{alpha2}.

The  procedure that we propose starts with an \emph{initial ansatz} for
the $n$-point functions to be used in the right hand side of the
flow equations.  Integrating the flow equation of a given
$n$-point function gives then the {\it leading order} (LO)
estimate for that $n$-point function. Inserting the leading order
of the $n$-point functions thus obtained in the right hand side of
the flow equations and integrating gives then the {\it
next-to-leading order} (NLO) estimate of the $n$-point functions.
And so on.

The equations will be solved starting at the bottom of the hierarchy, that is,
with the equation for the 2-point function. The flow equation for the 2-point
function involves in its right hand side the propagator (hence the 2-point function),
and the 4-point function. To determine the 2-point function in leading order, we need
therefore an initial ansatz for the propagator and the 4-point function. Similarly,
to get the 4-point function in leading order, we need an initial ansatz for the propagator,
for the 4-point function and the 6-point function. And so on.

There is no small parameter controling the convergence of the
process, and the terminology LO, NLO, refers merely to the number
of iterations involved in the calculation of the $n$-point
function considered. Obviously, the calculations become
increasingly complicated as the number of iterations increases,
and it is essential that the initial ansatz be as close as possible
to the exact solution so that only one or two iterations suffice
to get an accurate result. Our main task then is to
 construct such a good initial ansatz.

\subsection{The construction of the initial ansatz --- Generalities}

\label{initial-ansatz}
The initial ansatz for the $n$-point functions are the solutions of  approximate flow equations obtained by
making the following three approximations.

{\bf 1) Vertices are slowly varying functions of the external
momenta}

Our first approximation  (${\cal A}_1$) exploits a crucial property of the
NPRG: the derivative $\del_\kappa R_\kappa(q)$
limits the range of integrations in the flow equations to $q\simle
\kappa$. The momentum $q$ enters the vertices in the flow
equations typically in the form
$\Gamma_{12...n}^{(n)}(\kappa;p_1,p_2,...,p_{n-1}+q,p_n-q)$.
Approximation ${\cal A}_1$ consists then in assuming that, for any set of
external momenta $\lbrace p_1,p_2,...,p_n\rbrace$: \beq
\left|\frac{\Gamma_{12...n}^{(n)}(\kappa;p_1,p_2,...,p_{n-1}+q,p_n-q)-
\Gamma_{12...n}^{(n)}(\kappa;p_1,p_2,...,p_{n-1},p_n)}
{\Gamma_{12...n}^{(n)}(\kappa;p_1,p_2,...,p_{n-1},p_n)}\right| \ll
1 . \eeq This approximation is justified when the momenta $\lbrace
p_1,p_2,...,p_n\rbrace$ are much larger than  $\kappa$, since then
we can neglect $q$ compared to $p_i$, assuming that $\Gamma^{(n)}$ is a smooth function of the momenta when these are large. In the opposite case of
vanishing $p_i$'s,  we use the fact that the regulator insures
that  $\Gamma^{(n)}$  remains a smooth functions of its
arguments; in this case,  the approximation  ${\cal A}_1$ is analogous to the
leading order in the derivative expansion, i.e., to the LPA, known
to be a good approximation.

The approximation ${\cal A}_1$ is used to set
$q=0$ in the vertices $\Gamma^{(n)}$ and to factor them out of the
integrals in the r.h.s. of the flow equations.

{\bf 2) Propagators}

The second approximation (${\cal A}_2$) concerns the propagators in the
flow equation, for which we make the replacements:
\beq\label{appA2} G(p+q)\longrightarrow
G_{LPA'}(q)\,\Theta\left(1-\frac{\alpha^2 p^2}{\kappa^2} \right)
\eeq where $\alpha$ is an adjustable  parameter. A motivation for this approximation is the decoupling of high momentum modes in the flow equations, as illustrated in sect.~\ref{largeN}, and the parameter $\alpha$ will play here a role similar  to that it plays in sect.~\ref{largeN}. A measure of the quality of this approximation
is provided by fig.~\ref{fig:theta} which shows the ratio
$J_3^{(3)}(\kappa;p)/J_3^{(3)}(\kappa;p=0)$ where $J_d^{(n)}(\kappa;p)$ is defined in  eq.~(\ref{Jnk} ).
$J_3^{(3)}(\kappa;p)$ is the integral wich remains in
eq.~(\ref{4point}) after approximation ${\cal A}_1$ and after chosing as
 propagator that of the LPA' (the consistency of this choice
will be verified shortly). As seen in fig.~\ref{fig:theta},
$J_3^{(3)}(\kappa;p)/J_3^{(3)}(\kappa;p=0)$, as a function of
$p^2/\kappa^2$, looks indeed like a step function, with a weak
residual $\kappa$ dependence.

 Different criteria can be used to fix $\alpha$.
One may fix $\alpha$ so that the inflexion point of the curve in
fig.~\ref{fig:theta} is at $\alpha p=\kappa$. One then obtains,
for $N=2$, $\alpha\approx .9$. One can also adjust $\alpha$ so
that the integral over $\kappa$ of $J_3^{(3)}(\kappa;p)$ is identical to
that of $J_3^{(3)}(\kappa;0)\Theta(1-\alpha^2p^2/\kappa^2)$. This yields
$\alpha\approx .6$. We  regard these two possible choices as
extremes and we adopt the value $\alpha=.75\pm .15$ for our
leading order estimate in the case $N=2$.

Before moving to the next approximation, let us write the equations for the 2-point and 4-point
functions obtained at this stage, i.e., after approximations ${\cal A}_1$ and ${\cal A}_2$.  The equation for the 2-point function becomes
\begin{equation}
\label{self0}
\hspace{-1cm}\kappa\partial_\kappa
\Gamma^{(2)}_{12}(\kappa;p)= -\frac{1}{2}\Gamma^{(4)}_{12ll}(\kappa;p,-p,0,0) I_d^{(2)}(\kappa)
\, ,
\end{equation}
and that for the 4-point function reads
\begin{eqnarray}\label{gamma4A1A2}
&&\kappa\partial_\kappa\Gamma^{(4)}_{1234}(p_1,p_2,p_3,p_4)=
I_d^{(3)}(\kappa) \nonumber \\
&&\hspace{.5cm}\times\left\lbrace \Theta\left(\kappa^2-\alpha^2
{(p_1+p_2)^2}\right)
\;\Gamma^{(4)}_{12ij}(p_1,p_2,0,-p_1-p_2)\Gamma^{(4)}_{34ij}
(p_3,p_4,0,-p_3-p_4) \right.\nonumber \\
&&\hspace{.5cm}+\;\;\;\Theta\left({\kappa^2}-\alpha^2
{(p_1+p_3)^2}\right)
\;\Gamma^{(4)}_{13ij}(p_1,p_3,0,-p_1-p_3)
\Gamma^{(4)}_{24ij}(p_2,p_4,0,-p_2-p_4) \nonumber \\
&&\hspace{.5cm}\left.+\;\;\;\Theta\left(\kappa^2-\alpha^2
{(p_1+p_4)^2}
\right)\;\Gamma^{(4)}_{14ij}(p_1,p_4,0,-p_1-p_4)
\Gamma^{(4)}_{32ij}(p_3,p_2,0,-p_3-p_2) \right\rbrace \nonumber \\
&&\hspace{.5cm}-\frac{1}{2}\Gamma^{(6)}_{1234ii}(p_1,p_2,p_3,p_4,0,0)
\, I_d^{(2)}(\kappa), \end{eqnarray}
where the function $I_d^{(n)}(\kappa)$ is defined in eq.~(\ref{Ink}).

Note that the approximation ${\cal A}_2$ amounts to truncate severely the high
momentum tails of the propagators. This will cause inaccuracy at high
momenta, and a dependence of the leading order results on the
value of $\alpha$.

{\bf 3) Approximation for the (n+2)-point function}

In order to close the equation for $\Gamma^{(n)}$ we need an
approximation for $\Gamma^{(n+2)}$. Namely, we need an approximation for $\Gamma^{(4)}$ in the
equation for $\Gamma^{(2)}$ and for $\Gamma^{(6)}$ in the equation
for $\Gamma^{(4)}$.  Note that we do not want to perform a truncation of the hierarchy, as often done,  by setting to zero the higher order $n$-point functions: indeed in the scaling regime the contributions of all vertices are of the same order of magnitude. Rather, we shall try to obtain a rough estimate for $\Gamma^{(n+2)}$, which is sufficient in order to  get the
initial ansatz for $\Gamma^{(n)}$  (this rough estimate is not to be confused with the initial ansatz for
$\Gamma^{(n+2)}$).

In order to construct this estimate for $\Gamma^{(n+2)}$ we rely on the LPA' and also use an approximation
inspired by the analysis of the correlation functions in the $N
\to \infty$ limit of the previous section (sect.~\ref{largeN}). We consider explicitly here the equations for the 2-point and the 4-point functions.

In the case of the equation for $\Gamma^{(2)}$, one needs an approximation for
$\Gamma^{(4)}_{12ll}(\kappa;p,-p,0,0)$, as can be seen in eq.~(\ref{self0}). This can be  hinted  from
eq.~(\ref{4pointlargeN}), leading us to assume $\Gamma^{(4)}_{12ll}(\kappa;p,-p,0,0)=N\delta_{12} g_\kappa(0)$.
The resulting initial ansatz for $\Gamma^{(2)}$ is
simply a momentum independent function, the running mass, whose
flow equation is given by eq.~(\ref{deriveedemk2}). Therefore, our initial
ansatz for the propagator is consistent with eq.~(\ref{propLPA'}),  to within the small effect of the anomalous
dimension which is ignored at this stage.

 We turn now to $\Gamma^{(4)}$. As can be seen in  sect.~\ref{largeN}, after doing approximation  ${\cal A}_2$,
 the two types of terms in the right hand side of the flow
equation of $\Gamma^{(4)}$ are proportional, with a coefficient
that depends only on $\kappa$  (see eqs.~(\ref{B8}) and (\ref{B8b}), and fig.~ \ref{Fkp}). Our third approximation (${\cal A}_3$) consists in assuming that this property holds in general,
i.e., we set:
\beq\label{defFkA3}
\partial_\kappa \Gamma_{1234}^{(4)[6]}(p_1,p_2,p_3,p_4)=-F_\kappa\partial_\kappa
\Gamma_{1234}^{(4)[4]}(p_1,p_2,p_3,p_4) ,
\eeq
where in  the l.h.s. $\Gamma_{1234}^{(4)[6]}$
is the 6-point vertex contribution to the flow of $\Gamma^{(4)}$
 (last line in eq. (\ref{gamma4A1A2}) and fig.~\ref{fig:dGamma4_6}), while the term multiplying $-F_\kappa$
in the r.h.s is that including only 4-point vertices (the first
three lines in (\ref{gamma4A1A2}) and fig.~\ref{fig:4point}).
This relation becomes trivial  in the LPA, i.e., when all external momenta are zero. This allows us to fix
$F_\kappa$ from eq.~(\ref{defFk}).

Combining all approximations, one gets the following equation that needs to be solved in order to get the initial ansatz for $\Gamma^{(4)}$:
\begin{eqnarray}
\label{gamma40new}\small
&&\hspace{-0.6cm}\kappa\partial_\kappa\Gamma^{(4)}_{1234}(\kappa;p_1,p_2,p_3,p_4)=I_d^{(3)}(\kappa)\,
(1-F_\kappa)
  \nonumber \\
&&\hspace{.5cm}\times\left\lbrace \Theta\left({\kappa^2}-\alpha^2
{(p_1+p_2)^2}\right)
\Gamma^{(4)}_{12ij}(p_1,p_2,0,-p_1-p_2)\Gamma^{(4)}_{34ij}
(p_3,p_4,0,-p_3-p_4) \right.\nonumber \\
&&\hspace{.5cm}+\;\;\;\Theta\left({\kappa^2}-\alpha^2
{(p_1+p_3)^2}\right)
\Gamma^{(4)}_{13ij}(p_1,p_3,0,-p_1-p_3)
\Gamma^{(4)}_{24ij}(p_2,p_4,0,-p_2-p_4) \nonumber \\
&&\hspace{.5cm}\left.+\;\;\;\Theta\left({\kappa^2}-\alpha^2
{(p_1+p_4)^2}
\right)\Gamma^{(4)}_{14ij}(p_1,p_4,0,-p_1-p_4)
\Gamma^{(4)}_{32ij}(p_3,p_2,0,-p_3-p_2) \right\rbrace .\nonumber\\
\end{eqnarray}\normalsize

In the rest of this section, we construct the solution of this equation  in terms of the solution of the LPA'. As a simple illustration  of the method to be used, consider first the totally symmetric configuration of momenta:
$(p_1+p_2)^2=(p_1+p_3)^2=(p_1+p_4)^2=p^2$ (and
$p_1^2=p_2^2=p_3^2=p_4^2=3p^2/4$). One then distinguishes  in
eq.~(\ref{gamma40new}) two   regions, according to the value of
$\kappa$ relative to $\alpha p$. When $\kappa \geq \alpha p$, all the terms in eq.~(\ref{gamma40new}) are non-zero.
One can then verify that the LPA' expression of $\Gamma^{(4)}$, i.e., that given in eq.~(\ref{gamma4LPA}),
is a solution of the equation. Since the initial condition at $\kappa=\Lambda$ has the form
of eq.~(\ref{gamma4LPA}), and since eq.~(\ref{gamma40new}) is a first order differential equation in $\kappa$,
the LPA' solution is the unique solution for $\kappa \geq \alpha p$.
When $\kappa< \alpha p$, the r.h.s. of eq.~(\ref{gamma40new}) vanishes and the flow stops. In this region,
the solution remains the LPA' solution, but taken at the fixed value $\kappa=\alpha p$. These are the features that we uncovered when we analyzed the correlation functions in the large $N$ limit in sect.~\ref{largeN}.

A similar separation into different regions, according to the value of $\kappa$,
can be done for general momentum configurations.
In all cases, when $\kappa$ is larger than all the combinations of momenta appearing in
the $\Theta$-functions in eq.~(\ref{gamma40new}), the solution is simply the LPA' solution.
The other regions, where some of the $\Theta$-functions vanish, have to be analyzed case by case.
One can then solve eq.~(\ref{gamma40new}) in two steps:
first, for one vanishing momentum, $p_3=0$ and
$p_4=-p_1-p_2$, then for any combination.

In the next two subsections, in order to simplify the notation,
and  except when ambiguities may arise, we shall often omit to
indicate the explicit $\kappa$ dependence of $\Gamma^{(4)}$.

\subsection{Calculation of $\Gamma^{(4)}_{1234}(p_1,p_2,0,-p_1-p_2)$}
\label{gamma4p1p2}

In this case, eq.~(\ref{gamma40new}) reads:
\begin{eqnarray}\label{gamma41par}
&&\kappa \partial_\kappa \Gamma^{(4)}_{1234}(p_1,p_2,0,-p_1-p_2)=I_d^{(3)}(\kappa)(1-F_\kappa)\nonumber \\
&&\hspace{.5cm}\times\left\lbrace \Theta\left({\kappa^2}-\alpha^2
{(p_1+p_2)^2}\right)
\Gamma^{(4)}_{12ij}(p_1,p_2,0,-p_1-p_2)\Gamma^{(4)}_{34ij}(0,-p_1-p_2,0,p_1+p_2) \right.\nonumber \\
&&\hspace{.5cm}+\;\;\;\Theta\left({\kappa^2}-\alpha^2
{p_1^2}\right)\;\;\Gamma^{(4)}_{13ij}(p_1,0,0,-p_1)
\Gamma^{(4)}_{24ij}(p_2,-p_1-p_2,0,p_1) \nonumber \\
&&\hspace{.5cm}\left.+\;\;\;\Theta\left({\kappa^2}-\alpha^2
{p_2^2}\right)\;\;\Gamma^{(4)}_{14ij}(p_1,-p_1-p_2,0,p_2)
\Gamma^{(4)}_{32ij}(0,p_2,0,-p_2) \right\rbrace.
\end{eqnarray}
Notice that in each term in the r.h.s. there is one vertex
evaluated with two vanishing momenta. Furthermore, because of the
theta functions, each term gives a non-zero contribution only when
the remaining non vanishing momentum is smaller than
$\kappa/\alpha$. We are therefore in the conditions discussed at
the end of the last sub-section: the  4-point functions with two
vanishing momenta are simply the LPA' ones
(eq.~(\ref{gamma4LPA})). By using the fact that bosonic vertex
functions are completely symmetric under simultaneous exchange of
internal indices and momenta, we can rewrite
eq.~(\ref{gamma41par}) in the following way:
\begin{eqnarray}\label{geneqp1p2}
&&\kappa\partial_\kappa\Gamma^{(4)}_{1234}(p_1,p_2,0,-p_1-p_2)=g_\kappa I_d^{(3)}(\kappa)(1-F_\kappa)\nonumber \\
&&\hspace{.5cm}\times\left\lbrace \Theta\left({\kappa^2}-\alpha^2
{(p_1+p_2)^2}\right)
\left[\Gamma^{(4)}_{12ii}(p_1,p_2,0,-p_1-p_2)\delta_{34} \right.\right.\nonumber \\
&&\hspace{.5cm}\left.+\Gamma^{(4)}_{1234}(p_1,p_2,0,-p_1-p_2)
+\Gamma^{(4)}_{1243}(p_1,p_2,0,-p_1-p_2)\right]\nonumber \\
&&\hspace{.5cm}+\Theta\left({\kappa^2}-\alpha^2
{p_1^2}\right)
\left[\Gamma^{(4)}_{i2i4}(p_1,p_2,0,-p_1-p_2)\delta_{13} \right.\nonumber \\
&&\hspace{.5cm}\left.+\Gamma^{(4)}_{3214}(p_1,p_2,0,-p_1-p_2)
+\Gamma^{(4)}_{1234}(p_1,p_2,0,-p_1-p_2)\right]\nonumber \\
&&\hspace{.5cm}+\Theta\left({\kappa^2}-\alpha^2
{p_2^2}\right)
\left[\Gamma^{(4)}_{1ii4}(p_1,p_2,0,-p_1-p_2)\delta_{23} \right.\nonumber \\
&&\hspace{.5cm}\left.\left.+\Gamma^{(4)}_{1324}(p_1,p_2,0,-p_1-p_2)
+\Gamma^{(4)}_{1234}(p_1,p_2,0,-p_1-p_2)\right]\right\rbrace
\end{eqnarray}
This is a first order linear equation where the momenta are
parameters. To solve it, we can assume without loss of generality
that $p_1^2\ge p_2^2\ge (p_1+p_2)^2$. For the rest of this
section, except when that would lead to confusion, we will drop
the arguments of $\Gamma_{1234}^{(4)}$, being understood that
$\Gamma_{1234}^{(4)}$ refers to
$\Gamma_{1234}^{(4)}(\kappa;p_1,p_2,0,-p_1-p_2)$. We need to
consider four different regions, according to the value of
$\kappa$:

{\bf a) $\kappa \ge \alpha p_1$}. In this region,
the solution is identical to that of the LPA'.

{\bf b) $\alpha p_1 > \kappa \ge \alpha p_2$}. In this region,
eq.~(\ref{geneqp1p2}) becomes:
\begin{eqnarray}\label{eqf1}
&& \kappa\partial_\kappa\Gamma^{(4)}_{1234}=g_\kappa
I_d^{(3)}(\kappa)(1-F_\kappa)\left\lbrace
\Gamma^{(4)}_{12ii}\delta_{34} +2\Gamma^{(4)}_{1234}+\Gamma^{(4)}_{1243}
+\Gamma^{(4)}_{1ii4}\delta_{23}+\Gamma^{(4)}_{1324} \right\rbrace.
\end{eqnarray}
To solve this equation, we first notice that the solution is
symmetric under the exchange of the 2nd and the 4th internal
indices (with no exchange of the momenta), i.e.: \beq
\Gamma^{(4)}_{1234}(p_1,p_2,0,-p_1-p_2)=\Gamma^{(4)}_{1432}(p_1,p_2,0,-p_1-p_2)
. \eeq This property is true for $\kappa=\alpha p_1$, and one can
verify that it is maintained along the flow. We then look for the
general solution symmetric in the indices 2 and 4, in the form:
\beq\label{tendec1}
\Gamma^{(4)}_{1234}=(\delta_{12}\delta_{34}+\delta_{14}\delta_{32})\Gamma_A+\delta_{13}\delta_{24}\Gamma_B
. \eeq Substituing (\ref{tendec1}) in eq.~(\ref{eqf1}), one finds
the following system of linear equations: \beq\label{eqset1}
\left\lbrace
\begin{array}{c c l}
\kappa\partial_\kappa\Gamma_A &=& I_d^{(3)}(\kappa) g_\kappa  (1-F_\kappa)\left((N+4)\Gamma_A+2\Gamma_B\right) \\
\kappa\partial_\kappa\Gamma_B &=& I_d^{(3)}(\kappa) g_\kappa  (1-F_\kappa)\left(2\Gamma_A+2\Gamma_B\right) . \\
\end{array}
\right.
\eeq
The matrix
\begin{eqnarray}
\left(
\begin{array}{c c}
N+4 & 2 \\
2 & 2 \\
\end{array}
\right)
\end{eqnarray}
has the eigenvalues (which are both positives for $N>-2$) \beq
\lambda_{\pm}=\frac{N+6\pm\sqrt{N^2+4N+20}}{2} , \eeq
corresponding to the following eigenvectors: \beq
\left(\begin{array}{c}
\Gamma_A^{\pm}\\
\Gamma_B^{\pm}
\end{array}
\right) = \left(\begin{array}{c}
\frac{\lambda_{\pm}}{2}-1\\
1
\end{array}
\right) . \eeq Using these eigen-vectors, one can write the
general solution of eq.~(\ref{eqset1}) as:
\beq
\left(\begin{array}{c}
\Gamma_A\\
\Gamma_B
\end{array}
\right) =a_\kappa^+\left(\begin{array}{c}
\Gamma_A^+\\
\Gamma_B^+
\end{array}
\right) +a_\kappa^-\left(\begin{array}{c}
\Gamma_A^-\\
\Gamma_B^-
\end{array}
\right) ,
\eeq
where $a_\kappa^\pm$ verify:
\beq
\kappa\partial_\kappa a_\kappa^{\pm} = I^{(3)}_d(\kappa) g_\kappa
(1-F_\kappa)\lambda_{\pm}a_\kappa^{\pm}= \frac{\lambda_{\pm}}{N+8}
\kappa\partial_\kappa (\log g_\kappa) a_\kappa^{\pm}.
\eeq
We used
eq.~(\ref{gderivtotale}) to obtain this result. The equation above
can be integrated analytically, to give:
\beq
a_\kappa^{\pm}=a_{\alpha p_1}^{\pm}\left(\frac{g_\kappa}{g_{\alpha
p_1}}\right)^{\frac{\lambda_{\pm}}{N+8}} .
\eeq
By imposing
continuity between the two regions (a) and (b)  at $\kappa=\alpha
p_1$, we obtain then the solution in the region $\alpha p_1 >
\kappa \ge \alpha p_2$:
\begin{eqnarray}
&&\Gamma_{1234}^{(4)}=\frac{g_{\alpha p_1}}{\lambda_--\lambda_+}
\left\lbrace
\delta_{13}\delta_{24}\left[(\lambda_--4)\left(\frac{g_\kappa}{g_{\alpha
p_1}}\right)^{\frac{\lambda_+}{N+8}} \right.
-(\lambda_+-4)\left(\frac{g_\kappa}{g_{\alpha p_1}}\right)^{\frac{\lambda_-}{N+8}}\right] \nonumber \\
&& +(\delta_{12}\delta_{34}+\delta_{14}\delta_{23})
\left[-\lambda_+\left(\frac{g_\kappa}{g_{\alpha
p_1}}\right)^{\frac{\lambda_+}{N+8}}
\left.+\lambda_-\left(\frac{g_\kappa}{g_{\alpha
p_1}}\right)^{\frac{\lambda_-}{N+8}}
\right]\right\rbrace . \nonumber \\
\end{eqnarray}

{\bf c) $\alpha p_2 > \kappa \ge \alpha |p_1+p_2|$}. In this
region, eq.~(\ref{geneqp1p2}) becomes:
\beq\label{eqfc}
\partial_\kappa \Gamma^{(4)}_{1234}= \frac{1}{N+8}\left\lbrace \Gamma^{(4)}_{12ii}\delta_{34}
+\Gamma^{(4)}_{1234}+\Gamma^{(4)}_{1243} \right\rbrace \partial_\kappa (\log g_\kappa). \eeq We
need now the general tensor decomposition: \beq
\Gamma^{(4)}_{1234}=\delta_{12}\delta_{34}\Gamma_A+(\delta_{13}\delta_{24}+\delta_{14}\delta_{32})\Gamma_B
+(\delta_{13}\delta_{24}-\delta_{14}\delta_{32})\Gamma_C . \eeq By
substituting in eq.~(\ref{eqfc}) we get: \ \beq
\left\lbrace\label{eqset2}
\begin{array}{c c l}
\partial_\kappa\Gamma_A &=& \frac{1}{N+8}\left((N+2)\Gamma_A+2\Gamma_B\right)\partial_\kappa (\log g_\kappa) \\
\partial_\kappa\Gamma_B &=& \frac{2}{N+8} \Gamma_B\,\partial_\kappa (\log g_\kappa) \\
\partial_\kappa\Gamma_C &=& 0 . \\
\end{array}
\right.
\eeq
The antisymmetric sector ($\Gamma_C$) is decoupled. In order to get the solution in the symmetric
sector ($\Gamma_A, \Gamma_B$), we diagonalize the matrix:
\begin{eqnarray}
\left(
\begin{array}{c c}
N+2 & 2 \\
0 & 2 \\
\end{array}
\right),
\end{eqnarray}
and get  the eigenvalues
\beq
\mu_+=N+2 \; , \;\; \mu_-=2
\eeq
corresponding to the eigenvectors:
\begin{eqnarray}
\left( \begin{array}{c}
1\\
0
\end{array}
\right), \; \; \left(
\begin{array}{c}
1\\
-N/2
\end{array}
\right) .
\end{eqnarray}
One can then write the general solution of the symmetric part of
eq.~(\ref{eqset2}) as \beq \left(\begin{array}{c}
\Gamma_A\\
\Gamma_B
\end{array}
\right) =b^+_\kappa\left(\begin{array}{c}
1\\
0
\end{array}
\right) +b^-_\kappa\left(\begin{array}{c}
1\\
-N/2
\end{array}
\right) ,
\eeq
where $b_\kappa^{\pm}$ verifies
\beq
\kappa\partial_\kappa b_\kappa^{\pm} = I_d^{(3)}(\kappa) g_\kappa
(1-F_\kappa)\; \mu_{\pm}\; b_\kappa^{\pm},
\eeq
which, using
eq.~(\ref{gderivtotale}), leads to \beq b_\kappa^\pm=b_{\alpha
p_2}^\pm \left(\frac{g_\kappa}{g_{\alpha
p_2}}\right)^{\frac{\mu_{\pm}}{N+8}} .
\eeq
Imposing continuity between the regions (b) and (c), at $\kappa=\alpha p_2$, one
obtains finally:
\begin{eqnarray}
\label{resultatABC} \Gamma^{(4)}_{1234}&=&\delta_{12}\delta_{34}
\left(b_{\alpha p_2}^+ \left(\frac{g_\kappa}{g_{\alpha
p_2}}\right)^{\frac{\mu_+}{N+8}}
+b^-_{\alpha p_2}\left(\frac{g_\kappa}{g_{\alpha p_2}}\right)^{\frac{\mu_-}{N+8}}\right) \nonumber \\
&-&(\delta_{13}\delta_{24}+\delta_{14}\delta_{32})\frac{N}{2}
b_{\alpha p_2}^-\left(\frac{g_\kappa}{g_{\alpha p_2}}\right)^{\frac{\mu_-}{N+8}} \nonumber \\
&+&(\delta_{13}\delta_{24}-\delta_{14}\delta_{32})\Gamma^C_{\alpha
p_2} ,
\end{eqnarray}
where
\begin{eqnarray}
\label{a12p2}
&&b^+_{\alpha p_2}=\frac{g_{\alpha
p_1}}{\lambda_--\lambda_+}\left[\left(-\lambda_+ +\frac{1}{N}
(\lambda_--\lambda_+-4) \right)
\left(\frac{g_{\alpha p_2}}{g_{\alpha p_1}}\right)^{\frac{\lambda_+}{N+8}} \right.\nonumber \\
&&\hspace{1cm}\left. - \left(-\lambda_- +\frac{1}{N}
(\lambda_+-\lambda_--4) \right)
\left(\frac{g_{\alpha p_2}}{g_{\alpha p_1}}\right)^{\frac{\lambda_-}{N+8}} \right]  , \nonumber \\
&&b^-_{\alpha p_2}= -\frac{1}{N}\frac{g_{\alpha
p_1}}{\lambda_--\lambda_+}\left[(\lambda_--\lambda_+-4)
\left(\frac{g_{\alpha p_2}}{g_{\alpha
p_1}}\right)^{\frac{\lambda_+}{N+8}} \right.
\nonumber \\
&&\hspace{1cm}\left.- (\lambda_+-\lambda_--4) \left(\frac{g_{\alpha
p_2}}{g_{\alpha p_1}}\right)^{\frac{\lambda_-}{N+8}}\right] ,
\end{eqnarray}
\beq\label{gammaC} \Gamma_{\alpha p_2}^C = \frac{g_{\alpha
p_1}}{\lambda_--\lambda_+}\frac{N+2}{2}\left[
\left(\frac{g_{\alpha p_2}}{g_{\alpha
p_1}}\right)^{\frac{\lambda_+}{N+8}}- \left(\frac{g_{\alpha
p_2}}{g_{\alpha p_1}}\right)^{\frac{\lambda_-}{N+8}} \right] .
\eeq

{\bf d) $\alpha |p_1+p_2| > \kappa$}. In this region the flow
simply stops. The result is then:
\beq
\Gamma_{1234}^{(4)}(\kappa;p_1,p_2,0,-p_1-p_2)=\Gamma_{1234}^{(4)}(\kappa=\alpha|p_1+p_2|;p_1,p_2,0,-p_1-p_2)
\eeq

\subsection{Calculation of $\Gamma^{(4)}_{12ii}(p,-p,q,-q)$}

\label{gamma4pq}

At this point, we could solve eq.~(\ref{gamma40new}) for any
combination of momenta, given the fact that once the function is
known for the particular combination that has been treated in the
last section, all the information appearing in the right-hand side
of the equation is known. To give an example we shall consider in
this sub-section the explicit calculation of
$\Gamma^{(4)}_{12ii}(\kappa;p,-p,q,-q)$, when $q\leq \kappa$. The
result will be used in
the next section  as the initial
ansatz for $\Gamma^{(4)}$ in the calculation of the LO expression
of the self-energy. Note that since the favored values of $\alpha$
are smaller than 1, $\alpha q \leq \kappa$.

For the considered values of momenta, eq.~(\ref{gamma40new})
becomes:
\begin{eqnarray}\label{gamma4sundt}
&& \kappa\partial_\kappa \Gamma^{(4)}_{12ll}(p,-p,q,-q)=
I_d^{(3)}(\kappa)(1-F_\kappa)\left\lbrace
\Gamma^{(4)}_{12ij}(p,-p,0,0)
\Gamma^{(4)}_{llij}(q,-q,0,0) \right.\nonumber \\
&&\hspace{.5cm}+\Theta\left(1-\alpha^2\frac{(p+q)^2}{\kappa^2}\right)\Gamma^{(4)}_{1lij}(p,q,0,-p-q)
\Gamma^{(4)}_{2lij}(-p,-q,0,p+q) \nonumber \\
&&\hspace{.5cm}\left.+\Theta\left(1-\alpha^2\frac{(p-q)^2}{\kappa^2}\right)\Gamma^{(4)}_{1lij}(p,-q,0,-p+q)
\Gamma^{(4)}_{2lij}(-p,q,0,p-q) \right\rbrace .
\end{eqnarray}
The r.h.s. of this equation includes the expressions of
$\Gamma^{(4)}$ that have been determined in the previous
sub-section. It is useful to separate the contribution to
$\Gamma^{(4)}(\kappa;p,-p,q,-q)$ coming from the first line from
those coming from the second and third lines of
eq.~(\ref{gamma4sundt}) above. The first contribution corresponds
to the $s$-channel, the second corresponds to the sum of the $t$
and $u$ channels  (see fig.~\ref{fig:dGamma4}).

{\bf 1) $s$-channel}

In the $s$-channel, there are two kinematical regions:

{\bf a) $\alpha p\leq \kappa$}. In this case, we have \beq
\kappa\partial_\kappa \Gamma_{12ll}^{[s]}(p,-p,q,-q)=I_d^{(3)}(\kappa)(1-F_\kappa)
g^2_\kappa (N+2)^2 \delta_{12} , \eeq whose solution is (see
eq.~(\ref{gderivtotale})):
 \beq
\Gamma_{12ll}^{[s]}(p,-p,q,-q)=\frac{(N+2)^2}{N+8} g_\kappa
\delta_{12} . \eeq

{\bf b) $\alpha p > \kappa$}. In this region, we have: \beq
\kappa\partial_\kappa
\Gamma_{12ll}^{[s]}(p,-p,q,-q)=I_d^{(3)}(\kappa) (1-F_\kappa)
(N+2)^2 \delta_{12}g_\kappa g_{\alpha p} \left(
\frac{g_\kappa}{g_{\alpha p}}\right)^{\frac{N+2}{N+8}} , \eeq whose
solution is: \beq\label{eqDT}
\Gamma_{12ll}^{[s]}(p,-p,q,-q)=g_{\alpha p}
\delta_{12}(N+2)\left\{\frac{N+2}{N+8} +\left(
\frac{g_\kappa}{g_{\alpha p}} \right)^{\frac{N+2}{N+8}}-1\right\} ,
\eeq where we used continuity between the two regions in order to
fix the integration constant.

{\bf 2) $t$ and $u$ channels}

Let us now turn to the contribution of the $t$ and $u$ channels in
eq.~(\ref{gamma4sundt}). Since the two channels  only differ in
the sign of $q$, we consider only the $t$-channel,  with
kinematical variable $|p+q|$. There are two situations to analyze.

{\bf A) $|p+q|>|p|$}. In this case, there are two kinematical
regions:

{\bf a) $\alpha |p+q| \leq \kappa$}. In this region the
contribution in eq.~(\ref{gamma40new}) reads:
\beq
\kappa\partial_\kappa \Gamma_{12ll}^{[t]}(p,-p,q,-q)=I_d^{(3)}(\kappa)
(1-F_\kappa) g^2_\kappa \; 3 (N+2) \delta_{12} . \eeq The solution is easily obtained by using eq.~(\ref{gderivtotale}) in order to eliminate $F_\kappa$. One gets
 \beq\label{solsun}
\Gamma_{12ll}^{(t)}(p,-p,q,-q)=3\frac{N+2}{N+8} g_\kappa
\delta_{12} . \eeq

{\bf b) $\alpha |p+q| > \kappa$}. In this region the flow stops
and we obtain: \beq\label{eqSUN}
\Gamma_{12ll}^{[t]}(p,-p,q,-q)=3\frac{N+2}{N+8} g_{\alpha |p+q|}
\delta_{12} . \eeq

{\bf B) $|p+q|\le p$}. In this case there are three kinematical
regions:

{\bf a) $\alpha p \leq \kappa$}. This region is identical to
$(Aa)$ above. The solution is given by eq.~(\ref{solsun}).

{\bf b) $\alpha p > \kappa \geq \alpha |p+q|$}. Here the
contribution to eq.~(\ref{gamma40new}) becomes :
\begin{eqnarray}
&&\kappa\partial_\kappa \Gamma_{12ll}^{[t]}(p,-p,q,-q)=I_d^{(3)}(\kappa)
(1-F_\kappa) \delta_{12}\frac{N+2}{N^2+4N+20}g^2_{\alpha p}
\nonumber \\
&&\times \left\{
\left(\frac{3}{2}N^2+6N+30+\frac{N+14}{2}\sqrt{N^2+4N+20}\right)\left(\frac{g_\kappa}{g_{\alpha
p}} \right)^{\frac{2\lambda_+}{N+8}}
\right.\nonumber \\
&&\left.
+\left(\frac{3}{2}N^2+6N+30-\frac{N+14}{2}\sqrt{N^2+4N+20}\right)\left(\frac{g_\kappa}{g_{\alpha
p}} \right)^{\frac{2\lambda_-}{N+8}}
\right\} \nonumber \\
\end{eqnarray}
and has as solution:
\begin{eqnarray}\label{caseBb}
&&\Gamma_{12ll}^{[t]}(p,-p,q,-q)=3\delta_{12}\frac{N+2}{N+8}g_{\alpha
p} +\delta_{12}\frac{N+2}{N^2+4N+20}g_{\alpha p}
\nonumber \\
&&\times \left\{
\frac{\frac{3}{2}N^2+6N+30+\frac{N+14}{2}\sqrt{N^2+4N+20}}{2\lambda_+-N-8}
\left[\left(\frac{g_\kappa}{g_{\alpha p}} \right)^{\frac{2\lambda_+-N-8}{N+8}}-1\right] \right. \nonumber \\
&&\left.+\frac{\frac{3}{2}N^2+6N+30-\frac{N+14}{2}\sqrt{N^2+4N+20}}{2\lambda_--N-8}
\left[\left(\frac{g_\kappa}{g_{\alpha p}}
\right)^{\frac{2\lambda_--N-8}{N+8}}-1\right]
\right\} , \nonumber \\
\end{eqnarray}
where, again, we have imposed continuity.

{\bf c) $ \kappa < \alpha |p+q|$}. In this regime the flow stops.
The solution is found by fixing $\kappa = \alpha |p+q|$ in
eq.~(\ref{caseBb}).

\section{Leading order results} 

\label{sec:LO}

The self-energy $\Sigma(\kappa;p)$ is obtained by integrating
eq.~(\ref{eq:dGamma2}) from the microscopic scale $\Lambda$ to the given
value of the parameter $\kappa$:
\beq
\label{self3}
\delta_{12} \Sigma(\kappa;p) =\delta_{12} \, r-\frac{1}{2}\int_\Lambda^\kappa
d\kappa'\int\frac{d^dq}{(2\pi)^d} G^2(\kappa';q)
\partial_{\kappa'}R_{\kappa'}(q)\Gamma_{12ii}^{(4)}(\kappa';p,-p,q,-q) ,
\eeq where we have used the boundary condition
$\Sigma(\kappa=\Lambda;p)=r$, with $r$ the bare mass. We shall be working in the critical regime, i.e. for a
vanishing physical mass. Thus  $r$ is supposed to be adjusted so
that $\Sigma(\kappa=0;p=0)=0$, that is:
\beq\label{baremasseqn}
\delta_{12} \, r=-\frac{1}{2}\int_0^\Lambda
d\kappa'\int\frac{d^dq}{(2\pi)^d} G^2(\kappa';q)
\partial_{\kappa'}R_{\kappa'}(q)\Gamma_{12ii}^{(4)}(\kappa';0,0,q,-q).
\eeq
One may use this equation to eliminate the explicit $r$-dependence in eq.~(\ref{self3})
\begin{eqnarray}
\label{selfLO1}
\delta_{12}\Sigma(\kappa;p) &=&
-\frac{1}{2}\int_{0}^{\kappa}d\kappa'\int\frac{d^dq}{(2\pi)^d}
G^2(\kappa';q)\partial_{\kappa'}R_{\kappa'}(q)
\Gamma^{(4)}_{12ii}(\kappa';p,-p,q,-q) \nonumber \\
&&\hspace{-1.5cm}+\frac{1}{2}\int_0^{\Lambda}d\kappa'\int\frac{d^dq}{(2\pi)^d}
G^2(\kappa';q)\partial_{\kappa'}R_{\kappa'}(q)
\left(\Gamma^{(4)}_{12ii}(\kappa';p,-p,q,-q)-\Gamma^{(4)}_{12ii}(\kappa';0,0,q,-q)
\right),  \nonumber \\
\end{eqnarray}
from which one immediately deduces the following expression for the physical self-energy  $\Sigma(p)\equiv
\Sigma(\kappa=0;p)$:
\beq
\label{selfLO1b}
\delta_{12}\Sigma(p) =\frac{1}{2}\int_0^{\Lambda}d\kappa'\int\frac{d^dq}{(2\pi)^d}
G^2(\kappa';q)\partial_{\kappa'}R_{\kappa'}(q)
\left(\Gamma^{(4)}_{12ii}(\kappa';p,-p,q,-q)-\Gamma^{(4)}_{12ii}(\kappa';0,0,q,-q)
\right). \nonumber \\
\eeq
This expression   automatically  satisfies the criticality condition at $\Sigma(p=0)=0$. But, of course,  it holds provided eq.~(\ref{baremasseqn}) holds.

In  the following subsections  we  study the self-energy at leading order (LO) of our approximation scheme. The leading order consists in using in the r.h.s. of eq.~(\ref{selfLO1}) the initial ansatz for  $\Gamma_{12ii}^{(4)}(\kappa';p,-p,q,-q)$ that has been derived in the previous section. Note that for this initial ansatz,    $\Gamma_{12ii}^{(4)}(\kappa';0,0,q,-q)$ is given by the LPA' expression, so that eq.~(\ref{baremasseqn}) is satisfied at LO with the value of $r$ obtained by solving the LPA'  (eq.~(\ref{baremasseqn})  for $r$ is then equivalent to  eq.~(\ref{preMk}), a self-consistent equation for the running mass $m_\kappa$ where the value of $r$ is adjusted so that, for a given value of the bare coupling, $m_\kappa=0$)).

\subsection{The self-energy at LO}
\label{sec-selfLO} 

As we just mentioned, in order to calculate $\Sigma_{LO}$, we 
use, as input in the r.h.s. of eq.~(\ref{selfLO1}), the initial
ansatz for both the propagator and the 4-point function. The initial ansatz for the propagator is needed only for
$q<\kappa$ and is taken to be the LPA' propagator (see
eq.~(\ref{propLPA'})):
\begin{equation}
\label{LPAprop}
G_{LPA'}^{-1}(\kappa;q<\kappa) = Z_\kappa q^2 + m^2_\kappa + R_\kappa (q) =
Z_\kappa \kappa^2 (1+\hat m_\kappa^2) .
\end{equation}
The initial ansatz for $\Gamma^{(4)}$ was determined in section
\ref{gamma4pq}. It depends only on $\kappa$, $p^2$, $q^2$ and the
angle $\theta$ between $p$ and $q$. By performing the
integrations over all  angles other than $\theta$, one gets:
\begin{eqnarray}\label{selfLO2}
\delta_{12} \Sigma_{LO}(\kappa;p)&=& - \frac{d-1}{4\pi}K_{d-1}
\int_{0}^{\kappa}
d\kappa'\frac{1}{Z_{\kappa'} {\kappa'}^2 (1+\hat m_{\kappa'}^2)^2} \int_{0}^{\kappa'}
q^{d-1}\; dq \left( 2+\eta_{\kappa'} (\frac{q^2}{\kappa'^2} -1) \right) \nonumber \\
&&\hspace{-1cm}\times\int_{0}^\pi d\theta\sin\theta  (1-\cos^2\theta)^{(d-3)/2}
 \Gamma_{12ii}^{(4)}(\kappa';p,-p,q,-q) \nonumber \\
&+&\frac{d-1}{4\pi}K_{d-1} \int_{0}^{\Lambda}
d\kappa'\frac{1}{Z_{\kappa'} {\kappa'}^2 (1+\hat m_{\kappa'}^2)^2} \int_{0}^{\kappa'}
q^{d-1}\; dq \left( 2+\eta_{\kappa'} (\frac{q^2}{\kappa'^2} -1) \right) \nonumber \\
&&\hspace{-1cm}\times\int_{0}^\pi d\theta\sin\theta (1-\cos^2\theta)^{(d-3)/2}
 \left( \Gamma_{12ii}^{(4)}(\kappa';p,-p,q,-q)
-\Gamma_{12ii}^{(4)}(\kappa';0,0,q,-q) \right) . \nonumber \\
\end{eqnarray}
The physical self-energy at LO  is then given by:
\begin{eqnarray}
\label{selfLO}
\delta_{12} \Sigma_{LO}(p)  &=&\frac{d-1}{4\pi}K_{d-1} \int_{0}^\Lambda
d\kappa\frac{1}{Z_{\kappa} {\kappa}^3 (1+\hat m_{\kappa}^2)^2} \int_{0}^{\kappa}
q^{d-1}\; dq \left( 2+\eta_{\kappa} (\frac{q^2}{\kappa^2} -1) \right)  \nonumber \\
&&\hspace{-1cm}\times\int_{0}^\pi d\theta\sin\theta (1-\cos^2\theta)^{(d-3)/2}
 \left( \Gamma_{12ii}^{(4)}(\kappa;p,-p,q,-q)
-\Gamma_{12ii}^{(4)}(\kappa;0,0,q,-q) \right) .\nonumber\\
\end{eqnarray}

This expression 
has interesting scaling properties that we shall
present for the case $d=3$ (most of the discussion extends to arbitrary dimensions, with the replacement of $u$ by $u^{1/(4-d)}$).

First, a simple analysis shows that $\Sigma_{LO}(p)$ in
eq.~(\ref{selfLO}) can be written in the form $\Sigma_{LO}(p) = u^2 \hat
\Sigma(p/u)$ where $\hat\Sigma$ is a dimensionless function.  To see that, we note that, as seen in  section
\ref{gamma4pq}, $\Gamma_{12ii}^{(4)}(\kappa;p,-p,q,-q)$ is  proportional to the LPA'
function $g_l$ where $l=\kappa$, $\alpha p$, or $\alpha|p+q|$. Now, as discussed in
section \ref{LPA}, in $d=3$, the dimensionless function $\hat g_l\sim l^{d-4} g_l$ only depends on
$l/u$ if $u/\Lambda$ is small enough.  It follows that $\Gamma_{12ii}^{(4)}(\kappa;p,-p,q,-q)=u\hat\Gamma_{12ii}^{(4)}(\kappa/u;p/u,-p/u,q/u,-q/u)$ where $\hat\Gamma_{12ii}^{(4)}$ is a dimensionless function. The result for $\Sigma_{LO}$ follows after noticing that the remaining dependence in $\Lambda$ sits in the upper limit of integration: since the integral converges, that dependence becomes negligible when $\Lambda/u\gg 1$.

\begin{figure}[t!]
\begin{center}
\includegraphics*[width=12cm]{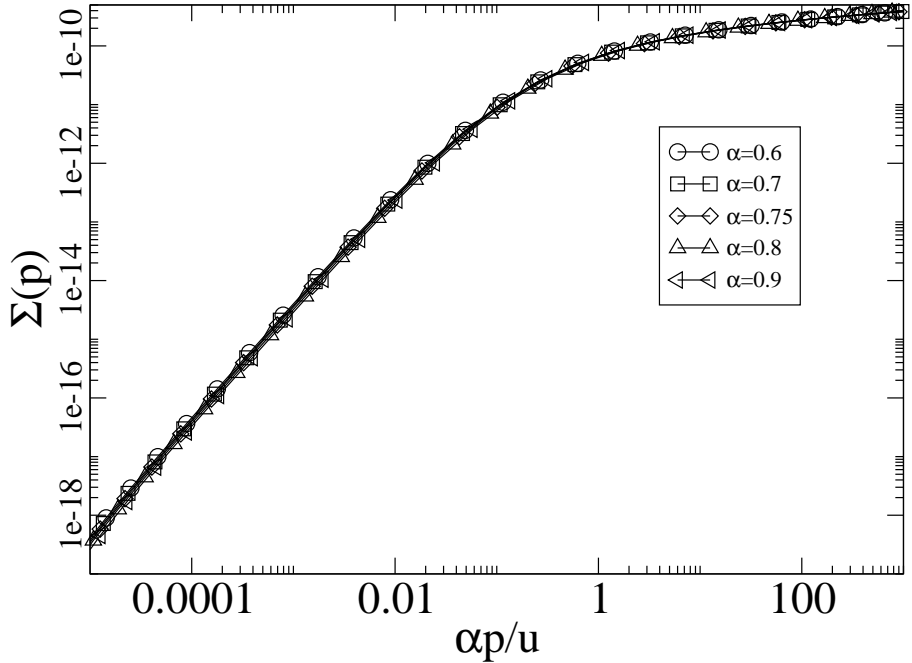}
\end{center}
\caption{\label{selfLO-alpha} $\Sigma(p)$ (in units of $\Lambda^2$)  as a function of $\alpha p/u$
for $N=2$ and various
values of $\alpha$: $\alpha=0.6$ (circles), $\alpha=0.7$ (square),
$\alpha=0.75$ (diamond), $\alpha=0.8$ (triangle up) and $\alpha=0.9$ (triangle left).
The curves exhibit the $\alpha$ scaling explained in the text.}
\end{figure}

A similar (but approximate) scaling holds for the dependence on
the parameter $\alpha$. To see that, let us set
 $q=0$ in the  4-point functions in eq.~(\ref{selfLO})
(similarly to what is done for  the approximation ${\cal A}_1$ of section
\ref{initial-ansatz}). Then, by using the explicit expressions of
$\Gamma^{(4)}(\kappa;p,-p,0,0)$ presented in sub-section
\ref{gamma4pq}, one  can verify  from eq.~(\ref{selfLO}) that
$\Sigma_{LO}(p)$ is a function of $\alpha p$ only, i.e.,
$\Sigma_{LO}(p)=\hat\Sigma(\alpha p)$. In fact, we expect this property
to be best satisfied for low values of $\alpha$: Indeed, since the
second line of eq.~(\ref{selfLO}) is non vanishing only for $q
\leq\kappa \leq \alpha p$ (see sub-section \ref{gamma4pq}), the
smaller the value of $\alpha$ the smaller the domain of variation
of $q$, and the better is the approximation $q=0$. The approximate scaling on $\alpha$ is clearly
visible in  fig.~\ref{selfLO-alpha}.

  Turning now to the momentum behavior of $\Sigma_{LO}$, we note  that 
both the low and high momentum regimes   are
correctly reproduced,  
independently of the value of $\alpha$. At large momenta,  we recover  the logarithmic behavior
predicted by second order perturbation theory, namely $\Sigma(p)\sim \ln (p/u)$. However, the numerical coefficient
in front of the logarithm (which does not depend on $\alpha$) comes
about 7\% higher than the correct one ($(N+2)u^2/(288 \pi^2)$). 

In the low momentum  region, we obtain the expected power law behavior 
$p^2+\Sigma_{LO}(p) \propto p^{2-\eta^*}$. It turns out that the exponent  $\eta^*$ is the
value of the function $\eta_\kappa$ at the IR fixed point of the LPA'. This is verified numerically with a numerical uncertainty of 0.001, independently of the value of the parameter $\alpha$. But this is also an exact result at the present level of approximation.  To prove this, let us
first note that, in eq.~(\ref{selfLO}), the difference of the two functions $\Gamma^{(4)}$   in the
second line is non-vanishing only when $\kappa \sim p$. Indeed, as can be easily
seen from their explicit expressions given  in section \ref{gamma4pq},
the two functions $\Gamma^{(4)}$ coincide when $\kappa>\alpha p$ and
$\kappa > \alpha |p\pm q|$; therefore there are two different situations where the contributions are not zero. The first situation is  $\kappa <\alpha p$. This implies that $\kappa<p$ (remember that $\alpha <1$). The second situation is more subtle. If we have $\kappa<\alpha |p\pm q|$, one has $|p|\ge |p\pm q|-|q|\ge |p\pm q|-\kappa\ge \kappa(1/\alpha-1)$, where we used the triangular inequality and the fact that $q\le \kappa$. In both situations we found that, as announced, the integrand in eq.~(\ref{selfLO})  is non vanishing only  when
$\kappa < \beta p$, where $\beta$ is a number of order 1. It follows that if
$p$ is in the scaling region, i.e., if $p \ll p_c$ (with $p_c\simeq \kappa_c$), so are
all the momentum variables in the integrand of eq.~(\ref{selfLO}), i.e., $p$, $q$, $|p\pm q|$, and $\kappa$.  Then all the functions
appearing in the r.h.s. of eq.~(\ref{selfLO})  are in the scaling
regime, and their dependence on $\kappa$ is controlled by the IR fixed point:
\beq
\hat m_{\kappa}^2 \simeq \hat m^{*2} \; , \;\;
\hat g_{\kappa} \simeq \hat g^* \; , \;\;
\eta_\kappa \simeq \eta^* \; , \;\;
Z_{\kappa} \propto \kappa^{-\eta^*} \; ,
\eeq
where we used eq.~(\ref{defZk}). Then, from eq.~(\ref{adimcons}), we get
\beq
\label{gIR}
g_\kappa \propto \kappa^{4-d-2\eta^*} .
\eeq
At this point we perform the change of
variables $\kappa=px$ and $q=py$ in order to make explicit the $p$
dependence of $\Sigma(p)$ in eq.~(\ref{selfLO}): we collect  a
factor $p^{\eta^*}$ from $Z^{-1}_{\kappa}$ and an overall factor  $p^{d-2}$ due to the terms  $d\kappa$,
$dq$, $\kappa^{-3}$ and $q^{d-1}$  appearing in the integrand. As for the $p$ dependence
of the 4-point functions, one uses the fact that $\Gamma^{(4)}(\kappa;p,-p,q,-q)$ is proportional to $g_l f(g_{\kappa}/g_l)$ where $l$ is either $\alpha p$ or
$\alpha |p\pm q|$, and $f$ a dimensionless function (see section  \ref{gamma4pq}). After the change of variables,
using eq.~(\ref{gIR}), $\Gamma^{(4)}(\kappa;p,-p,q,-q)-\Gamma^{(4)}(\kappa;p,-p,0,0)$
can thus be written as $p^{4-d-2\eta^*}$ times a function of $x$ and $y$.
Altogether, and using the fact that, as shown above, $\Sigma_{LO}(p)=u^2\tilde \Sigma(p/u)$,
one gets:
\beq
\Sigma_{LO}(p) = {\cal C} u^{\eta^*}p^{2-\eta^*},
\eeq
where the proportionality coefficient ${\cal C}$ is the remaining dimensionless and finite
integral over $x$, $y$ (and $\theta$), which only depends on the parameter $\alpha$.

The anomalous dimension obtained from the present calculation is
then identical to that calculated  in the  LPA' (see eqs.~(\ref{equationpoureta})
and (\ref{defZk})). Its value, $\eta^* \sim 0.044$
 is   to be compared with  the best  estimates available
in the literature, e.g. $\eta=0.0354\pm0.0025$ \cite{refeta}.  A simple proof  that the dependence of the field renormalisation factor on the scale  $\kappa$ determines in general 
 the momentum dependence of the self-energy  (thus defining the anomalous dimension)
is presented in app.~\ref{appendixeta}. However there is no guarantee that this property should hold in any approximation (for instance it does not hold in the derivative expansion).  It is therefore gratifying to see that that the power law behavior expected for the momentum dependence of the self-energy in the scaling regime comes out naturally in the LO of  the present approximation scheme.

\subsection{Calculation of $\Delta\langle \varphi^2\rangle$}
\label{calcul_de_c}

As a further test of the quality of the leading order result for the self-energy, we have used $\Sigma_{LO}$ to calculate 
  the shift $\Delta T_c$ of the transition temperature of a dilute,  weakly
interacting, Bose gas. It has been shown that $\Delta T_c$
is linear in  $an^{1/3}$ \cite{club}, where $a$ is the scattering
length and $n$ the particle density:
\beq\label{deltaTc}
\frac{\Delta T_c}{T_c^0}=c \,\,a n^{1/3}.
\eeq
Here $T_c^0$ is the
condensation temperature of the ideal gas and $\Delta T_c=T_c-T_c^0$ with $T_c$  the
transition temperature of the interacting system. As shown in
Ref.~\cite{club},
  the coefficient $c$ can be related to the change $\Delta\langle
\varphi^2\rangle$ in the magnitude of the fluctuations of the field described by the
action (\ref{classicalaction}):   
\beq
c\,=-\frac{256 \pi^3}{\left(\zeta(3/2)\right)^{4/3}} \,
\frac{\Delta\langle\varphi_i^2\rangle}{Nu}, 
\eeq
in the limit  $u\to 0$ (and for $N=2$). 

The best numerical estimates for $\Delta\langle \varphi^2\rangle$,
and hence for $c$,  are those which have been obtained using the
lattice technique by two groups, with the results: $c=1.32\pm0.02$
\cite{latt2} and $c=1.29\pm 0.05$ \cite{latt1}. The
availabilty of these results has turned the calculation of $c$
into a testing ground for other non perturbative methods:
expansion in  $1/N$ \cite{BigN,Arnold:2000ef}, optimized
perturbation theory \cite{souza,Kneur04},
   resummed    perturbative  calculations to high loop orders
\cite{Kastening:2003iu}. Note that while the latter methods
yield critical exponents with several significant digits, they
predict $c$ with only a 10\% accuracy. This illustrates the
difficulty of getting an accurate determination of $c$ using
(semi) analytical techniques.

\begin{figure}[t!]
\begin{center}
\includegraphics*[width=12cm,angle=-90]{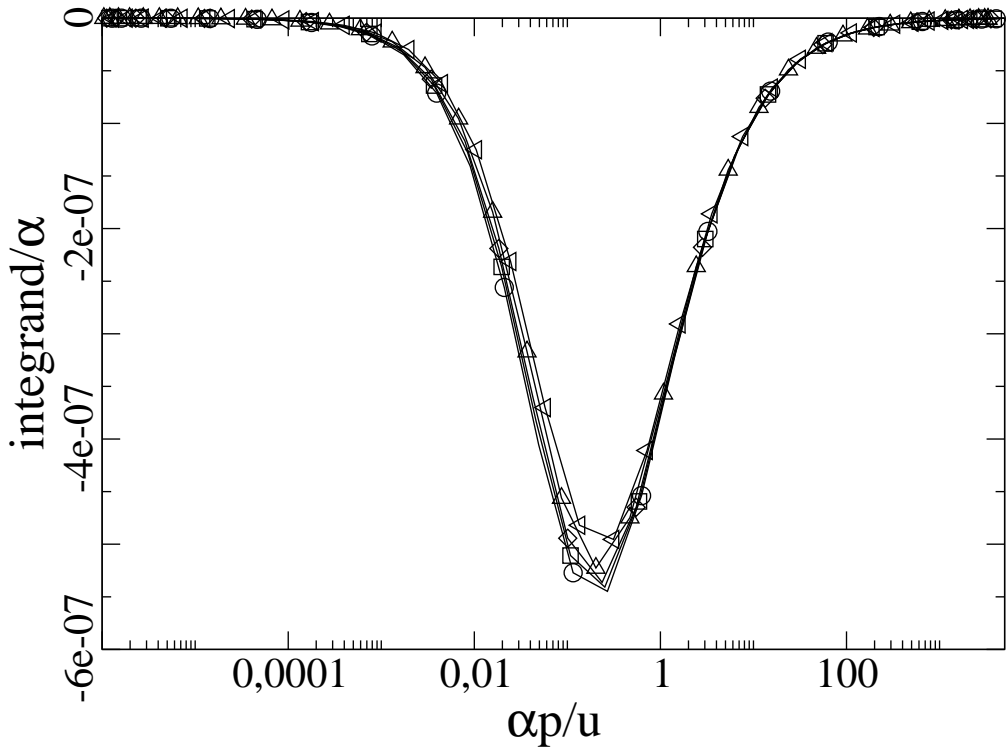}
\end{center}
\caption{\label{int-alpha} The integrand of eq.~(\ref{integralc})
(divided by $\alpha$, and in units of $\Lambda$) as a function of $\alpha p/u$
for various values of $\alpha$: $\alpha=0.6$ (circles), $\alpha=0.7$ (square),
$\alpha=0.75$ (diamond), $\alpha=0.8$ (triangle up) and $\alpha=0.9$ (triangle left)
(points shown are those needed to the numerical calculation of
the integral in eq.~(\ref{integralc}).) The curves exhibit the approximate  $\alpha$ scaling explained in the text.}
\end{figure}

To understand better the origin of the difficulty, let us write
$\Delta\langle \varphi_i^2\rangle$ as the following integral
\beq\label{integralc}
\frac{\Delta\langle \varphi_i^2\rangle}{N}=
\int\frac{\d^3 p}{(2\pi)^3}\,\left(
\frac{1}{p^2+\Sigma(p)}-\frac{1}{p^2}\right)= -\frac{1}{2\pi^2}\int\frac{dp}{p}\left[ p-\frac{p^3}{p^2+\Sigma(p)}\right].
\eeq
where
$\Sigma(p)$ is the self-energy at  criticality, i.e.,
$\Sigma(0)=0$. In eq.~(\ref{integralc}),  the term within the square brakets, to be referred below as the integrand, is, to a very good  approximation, equal to $\Sigma(p)/p$ (one finds numerically that this is a good approximation as soon as $p/u \simge 10^{-5}$). As we shall see shortly, $\Sigma(p)/p$ is peaked in the region of intermediate momenta between
the  critical region  and  the
high momentum perturbative region (see fig.~\ref{int-alpha}). 
The difficulty in getting a precise evaluation of the integral
(\ref{integralc}) is that it requires an  accurate determination
of $\Sigma(p)$ in a large region of momenta including the
crossover region between two different physical regimes
\cite{bigbec,BigN}. In that sense, the calculation of $c$ can be viewed as a very stringent test of the approximation scheme.

 A plot of the integrand of eq.~(\ref{integralc}) (divided by $\alpha$) is
shown in fig.~\ref{int-alpha}, for various values of $\alpha$. As announced, the momentum at which the integrand reaches its maximum lies in the intermediate momentum region:  in
fig.~\ref{int-alpha} this is  $\alpha p /u \simeq 0.2$.
The approximate scaling behaviour that  can be observed in fig.~\ref{int-alpha} follows from the property of the self-energy discussed in sect.~\ref{sec-selfLO} : as we have seen there, $\Sigma_{LO}(p;\alpha)\simeq\bar\Sigma(\alpha p)$, so that, setting $\bar p=\alpha p$
\beq\label{alphac}
\frac{\Delta\langle \varphi_i^2\rangle}{N}\approx - \alpha
\int\frac{\d^3 \bar p}{(2\pi)^3}\,
\frac{\bar \Sigma(\bar p)}{\bar p^4}.
\eeq

\begin{figure}[t!]
\begin{center}
\includegraphics*[width=10cm]{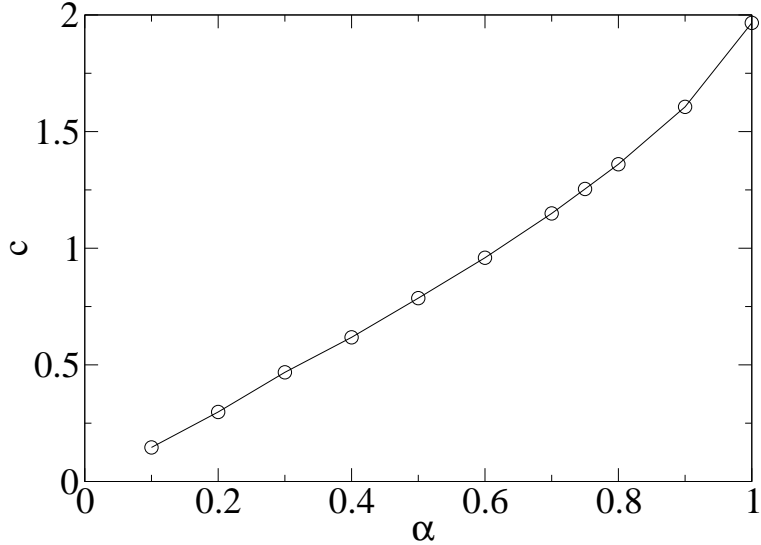}
\end{center}
\caption{\label{cLO} The coefficient $c$ calculated in L.O. as a function of the parameter $\alpha$.}
\end{figure}

Fig.~\ref{cLO} shows the value of the coefficient $c$ as a
function of  $\alpha$. The (almost) linear behavior of
$c$ as a function of $\alpha$
follows directly from eq.~(\ref{alphac}). Deviations from the non linear behaviour can be seen for $\alpha\simge  .7$: as we have discussed in the previous subsection,  for these large values of $\alpha$, the approximation
$\Sigma_{LO}(p;\alpha)\simeq\bar\Sigma(\alpha p)$ becomes less accurate.
 As we can see, when
$\alpha = 0.75\pm 0.15$, one gets $c= 1.3\pm 0.3$. This
result  confirms the quality of  the leading order expression of  the self-energy  for all values of the momentum.

We have also calculated $\Sigma_{LO}$  for different values of  $N$, and compared the corresponding results with those obtained by different means and  available in the literature. The quality of our numerical estimates remains of the same level as long as $N \simle 50$, but for larger values of $N$, the calculations lose accuracy. The range of acceptable values of $\alpha$ (see sect.~\ref{initial-ansatz}) remains the interval  $\sim 0.6-0.9$, and the resulting error bars on the predicted value of  $c$ stay of the order of $23-29\%$. One gets, for $N=1$, $c=1.06\pm 0.27$; for $N=3$, $c=1.47\pm 0.39$;
for $N=4$,  $c=1.66\pm 0.44$; for $N=10$, $c=2.33\pm 0.60$; for $N=40$, $c=2.97\pm 0.63$. These numbers are to be compared with those obtained using other methods; lattice calculation
 \cite{latt} or ressumed perturbation theory carried up to 7-loop order \cite{Kastening:2003iu} give: for $N=1$, $c=1.09\pm 0.09$ (lattice) and $c=1.07\pm
 0.10$ (7-loops); for $N=3$, $c=1.43\pm 0.11$ (7-loops); for $N=4$, 
 $c=1.60\pm 0.10$ (lattice) and $c=1.54\pm 0.11$ (7-loops). The exact result for $N\to \infty$ is also known 
 \cite{BigN}: $c=2.33$. One observes that, for all values of $N$, the best accepted results always lie within the
 error bars of our LO prediction; they approach the lower limit of the band when $N$ grows (the origin of the latter property can  in fact be understood by analyzing the steps leading to eq.~(\ref{B8b})).

\begin{figure}[t!]
\begin{center}
\includegraphics*[width=12 cm]{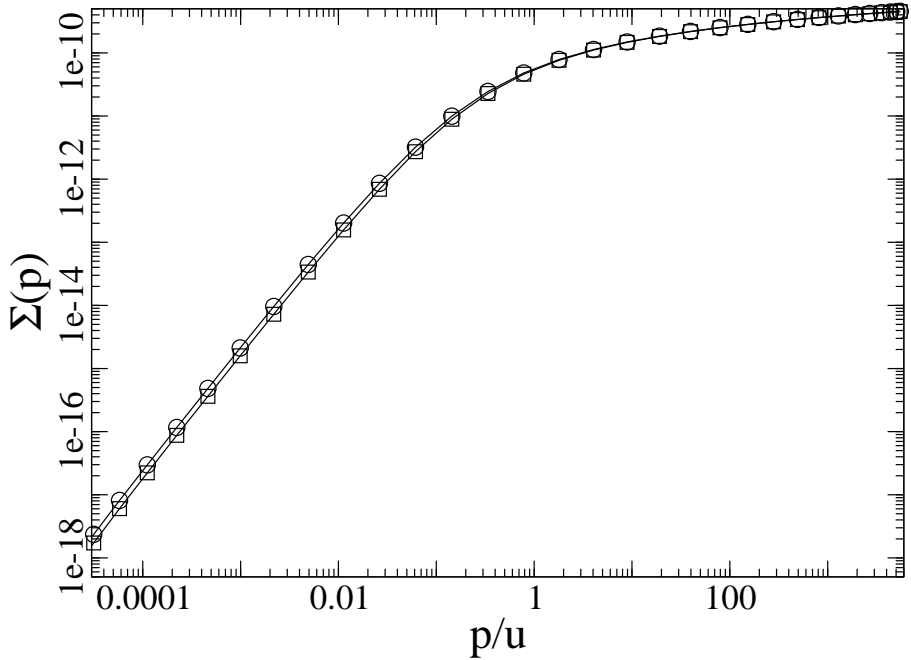}
\end{center}
\caption{\label{aprox1} Calculation of the self-energy (in units of $\Lambda^2$)  used to test approximation ${\cal A}_1$, as explained
in the text: the complete expression of $\Sigma(p)$ (triangles) and the
approximate one (squares). }
\end{figure}

Before finishing this section we  present a consistency check of the approximation
${\cal A}_1$  made in sect.~\ref{initial-ansatz} to construct the initial ansatz for the
4-point function $\Gamma^{(4)}$. This approximation consists in  neglecting
the internal momentum dependence in the 4-point vertices appearing in
the r.h.s. of the  flow equation. This was done in order to obtain eq.~(\ref{gamma4A1A2}) for $\Gamma^{(4)}$. Here we shall make this approximation
${\cal A}_1$  in
the equation for the self-energy, eq.~ (\ref{selfLO}).
Fig.~\ref{aprox1} compares the self-energies obtained form eq.~(\ref{selfLO}) with and without approximation ${\cal A}_1$. One can see that the approximate result differs very little from the exact one.
It turns out that both the perturbative regime and the exponent in the scaling regime 
are almost unchanged, most of the difference being concentrated in the intermediate momentum region. 
This is verified by calculating the coefficient $c$  with and without the approximation ${\cal A}_1$: The value obtained with  ${\cal A}_1$ is about 10\%
smaller than that obtained with $\Sigma_{LO}$. This illustrates the
large sensitivity of the coefficient $c$ to variations of the self-energy
in the cross-over region.

\section{Conclusions}

\label{conclusions}

The calculation of the self-energy of the O($N$) model demonstrates that the approximation scheme that we have presented fulfills its goal, that is, it offers a simple way to calculate the full momentum dependence of the $n$-point function. The accuracy achieved in  the leading order is already satisfactory, over the full momentum range, as shown by the various tests that we performed. 

A crucial ingredient in the calculation is the construction of the initial ansatz for the 4-point function. That in itself is an important part of the present paper. This initial ansatz is obtained by solving an approximate flow equation derived using well motivated approximations. The resulting 4-point function, albeit approximate, exhibits a realistic momentum dependence, also in the entire momentum range. In particular, the power law behavior expected in the scaling regime is reproduced.

The approximations that we have introduced to construct the initial ansatz for the 4-point function involve a parameter $\alpha$ that needs to be adjusted in such a way that approximate expressions match best the exact expressions that they are supposed to represent. This introduces a theoretical uncertainty, which, in the case of the calculation of the shift of the Bose-Eisntein transition temperature that we have presented, is of the order of 25\%. 

In a forthcoming paper \cite{alpha2}, we shall present   results  of a next-to-leading order analysis for the self-energy. To do so we shall need to   improve the accuracy of the 4-point function, as compared to the initial ansatz presented in this paper. That is, we shall calculate  the 4-point function at leading order, i.e., construct  an initial ansatz for the 6-point function. The next-to leading order calculation of the self-energy will allow us to test fully the approximation scheme, and detect some of its weaknesses. As we shall see, the calculation of $c$ will be greatly improved, in particular the dependence on the parameter $\alpha$ will be eliminated, and results obtained in quite good agreement with lattice data.

\appendix

\section {The function $\eta_\kappa$ and the anomalous dimension}

\label{appendixeta}

It is usually accepted \cite{Berges02} that the $\kappa$-dependence of
the field renormalisation factor $Z_\kappa$ (for values
of $\kappa$ where the couplings have approached the infrared fixed point) determines the anomalous dimension of the field.   However it is not a priori obvious
that the momentum dependence of correlation functions obtained after some approximation follows automatically the corresponding scaling law: this is not so for instance in the derivative expansion. We find it therefore useful to present  in this appendix a simple derivation of this property. The arguments in the proof given here help to clarify  the conditions  under which this property will be satisfied in an approximation. This also completes the derivation  presented in Sect.~\ref{sec-selfLO},  that the anomalous dimension calculated from the momentum dependence of the 4-function obtained as solution of the approximate equation derived in Sect.~\ref{Approximation_scheme} is indeed equal to that deduced from the $\kappa$-dependence of $Z_\kappa$ calculated in the LPA'.

Let us  consider
the 2-point function $\Gamma^{(2)}(p,\kappa,u)$ for $p, \kappa \ll \kappa_c\sim u^{1/(4-d)}$ in order to be in the scaling regime. Then, scale invariance implies that :
\beq
\frac{\Gamma^{(2)}(p',\kappa,u)}{\Gamma^{(2)}(p,\kappa,u)}=\tilde f\left(\frac{p'}{p},\frac{p}{\kappa}\right)
\eeq
where $\tilde f$ is a dimensionless function of its arguments. It follows that
\beq\label{Gamma2pu}
\Gamma^{(2)}(p,\kappa,u)=\Gamma^{(2)}(0,\kappa,u) \,f\left(\frac{p}{\kappa}\right)
\eeq
where we have set $f(p/\kappa)\equiv \tilde f(0,p/\kappa)$. Note that the dependence on the  microscopic parameter $u$ is entirely contained in the factor $\Gamma^{(2)}(0,\kappa,u)$ ($\Gamma^{(2)}(0,\kappa,u)$ is well defined thanks to the IR regulator), and the momentum dependence factors out in the scaling function $f(p/\kappa)$. This function has a Taylor expansion  at small $p/\kappa$, and $f(0)=1$.
At this point we may use scale invariance again, together with dimensional analysis, in order to show that in the regime $\kappa\ll \kappa_c$:
\beq\label{gammaeta}
\Gamma^{(2)}(0,\kappa,u)\propto\kappa^2 \left(\frac{\kappa}{u^{1/(4-d)}}\right)^{-\eta},
\eeq
where $\eta$ is constant.
It then follows that for the function $\Gamma^{(2)}(p,\kappa,u)$ in eq.~(\ref{Gamma2pu})  to have a
limit when $\kappa\to 0$, we must have, for large values of $p/\kappa$:
\beq
f\left(\frac{p}{\kappa}\right)\propto \left(\frac{p}{\kappa}\right)^{2-\eta}
\eeq
where $\eta$ is the same constant as in eq.~(\ref{gammaeta}). Thus,
\beq
\Gamma^{(2)}(p,\kappa,u)\propto p^2\times\left(\frac{p}{u^{1/(4-d)}}\right)^{-\eta}.
\eeq

We can write:
\beq
\Gamma^{(2)}(p,\kappa,u)-\Gamma^{(2)}(0,\kappa,u)=Z_\kappa p^2+O(p^4).
\eeq
and from eq.~(\ref{Gamma2pu})
\beq\label{expGamma2}
\Gamma^{(2)}(p,\kappa,u)-\Gamma^{(2)}(0,\kappa,u)=(f(p/\kappa)-1)\Gamma^{(2)}(0,\kappa,u).
\eeq
In the regime $\kappa\to 0$, so that eq.~(\ref{gammaeta}) is valid, and $p\ll \kappa$ so that $f(p/\kappa)\approx 1+C\; (p/\kappa)^2$, with $C$ a numerical constant, one can then use eqs.~(\ref{gammaeta}) and (\ref{expGamma2}) to deduce the behaviour of $Z_\kappa$:
\beq
Z_\kappa=C\left(\frac{\kappa}{u}\right)^{-\eta}.
\eeq

\section{The function $J_3^{(3)}(\kappa;p)$}

\label{functionJ}

 Using the LPA' propagator (see eq.~(\ref{LPAprop}))
and the  regulator of eq.~(\ref{defRk}), making the change of variables
$\bar p=p/\kappa, v=q/\kappa$ and $\cos \theta=p.q/p\;q$, and performing the integrals over
the remaining angular variables, one can write eq.~(\ref{Jnk}) as:
\begin{eqnarray}
J_d^{(3)}(\kappa;p)
\hspace{-.2cm}&=&\hspace{-.2cm}\frac{\kappa^{d-4}}{Z_\kappa^2(2\pi)^d}\frac{2\pi^{\frac{d-1}{2}}}
{\Gamma(\frac{d-1}{2})}\frac{1}{(1+\hat m_\kappa^2)^2}
\int_0^1 dv v^{d-1}\int_{0}^{\pi}d \theta \sin \theta
(1-\cos^2 \theta)^{\frac{d-3}{2}} \nonumber \\
\hspace{-.8cm}\times &&\hspace{-.8cm}\frac{(2-\eta+\eta v^2)}{\Theta(1-v^2-\bar p^2+2 v \bar p \cos \theta)+
(v^2+\bar p^2-2 v \bar p \cos \theta)
\Theta(v^2+\bar p^2-2 v \bar p \cos \theta-1)+\hat m_\kappa^2}. \nonumber \\
\end{eqnarray}
This expression is valid for arbitrary $d$, but we shall evaluate it only for $d=3$. In order to take care of the  $\Theta$ functions it is convenient
to separate the calculation in two different regions: $2<\bar p$ and $\bar p\leq 2$. In each case, one performs the $\theta$ integral first, and then the integral over  $v$. One  gets:

a) $2<\bar p$
\begin{eqnarray}
 &&\hspace{-.2cm}J_3^{(3)}(\kappa;p)
=\frac{1}{\kappa Z_\kappa^2(2\pi)^2}\frac{1}{(1+\hat m_\kappa^2)^2}\left\lbrace 2
+\frac{\eta}{2}\left(-\frac{5}{3}+\bar p^2-3\hat m_\kappa^2\right) \right.
\nonumber \\
&&+\frac{1}{2\bar p}\left[-1+\frac{\eta}{4}+\left(\bar p+\sqrt{-\hat m_\kappa^2}\right)^2
\left(1-\frac{\eta}{2}
+\frac{\eta}{4}\left(\bar p+\sqrt{-\hat m_\kappa^2}\right)^2\right)\right]
\log\left(\frac{\bar p-1+\sqrt{-\hat m_\kappa^2}}{\bar p+1+\sqrt{-\hat m_\kappa^2}}
\right) \nonumber \\
&&+\left.\frac{1}{2\bar p}\left[-1+\frac{\eta}{4}+\left(\bar p-\sqrt{-\hat m_\kappa^2}
\right)^2\left(1-\frac{\eta}{2}
+\frac{\eta}{4}\left(\bar p-\sqrt{-\hat m_\kappa^2}\right)^2\right)\right]
\log\left(\frac{\bar p-1-\sqrt{-\hat m_\kappa^2}}{\bar p+1-\sqrt{-\hat m_\kappa^2}}
\right) \right\rbrace \nonumber \\
&&=\frac{1}{\kappa  Z_\kappa^2(2\pi)^2}\frac{1}{(1+\hat m_\kappa^2)^2}\left\lbrace
\frac{4}{\bar p^2}\left(\frac{1}{3}-\frac{\eta}{15}\right)+\frac{4}{\bar p^4}\left(
\frac{1}{15}-\frac{\eta}{105}
-\frac{\hat m_\kappa^2}{3}+\frac{\eta\hat m_\kappa^2}{15}\right)+
\mathcal{O}(1/(\bar p^6)) \right\rbrace \nonumber \\
\end{eqnarray}

c) $\bar p\leq 2$.
\begin{eqnarray}
J_3^{(3)}(\kappa;p)
 \hspace{-.2cm}
&=&\hspace{-.2cm}\frac{\kappa^{-1}}{Z_\kappa^2(2\pi)^2(1+\hat m_\kappa^2)^2}\left\lbrace
-1+\frac{\eta}{4}+\frac{\eta\hat m_\kappa^2}{4}+\bar p\left(\frac{3}{2}
-\frac{\eta}{8}-\frac{7\eta\hat m_\kappa^2}{8}\right)
-\frac{3\eta}{4}\bar p^2\right. \nonumber \\
&+&\frac{25\eta}{48}\bar p^3+\frac{1}{1+\hat m_\kappa^2}\left(\frac{4}{3}
-\frac{4\eta}{15}-\bar p+\frac{\eta}{3}\bar p^2
+\left(\frac{1}{12}-\frac{\eta}{6}\right)\bar p^3+\frac{\eta}{120}\bar p^5\right)
\nonumber \\
&+&\hspace{-.2cm}\frac{1}{2\bar p}\left[1-\frac{\eta}{4}
-\left(\bar p+\sqrt{-\hat m_\kappa^2}\right)^2
\left(1-\frac{\eta}{2}+\frac{\eta}{4}
\left(\bar p+\sqrt{-\hat m_\kappa^2}\right)^2\right)\right]
\log \left(\frac{\bar p+1+\sqrt{-\hat m_\kappa^2}}{1+\sqrt{-\hat m_\kappa^2}}\right)
\nonumber \\
&+&\left.\hspace{-.2cm}\frac{1}{2\bar p}\left[1-\frac{\eta}{4}-
\left(\bar p-\sqrt{-\hat m_\kappa^2}\right)^2\left(1-\frac{\eta}{2}+\frac{\eta}{4}
\left(\bar p-\sqrt{-\hat m_\kappa^2}\right)^2\right)\right]
\log\left(\frac{\bar p+1-\sqrt{-\hat m_\kappa^2}}{1-\sqrt{-\hat m_\kappa^2}}\right)
\right\rbrace\nonumber \\
&=&\hspace{-.2cm}\frac{\kappa^{-1}}{Z_\kappa^2(2\pi)^2(1+\hat m_\kappa^2)^2}\left\lbrace
\frac{4}{3(1+\hat m_\kappa^2)}\left(1-\frac{\eta}{5}\right)-\frac{2}{3(1+\hat m_\kappa^2)^2}\bar p^2 \right.\nonumber \\
&+&\left. \frac{2+\eta-2\hat m_\kappa^2+\eta \hat m_\kappa^2}{6(1+\hat m_\kappa^2)^3}\bar p^3
-\frac{2(1+\eta-5\hat m_\kappa^2+\eta \hat m_\kappa^2)}{15(1+\hat m_\kappa^2)^4}\bar p^4+\mathcal{O}(\bar p^5)\right\rbrace.
\end{eqnarray}  
 
 \acknowledgements
Authors R. M-G and N. W are grateful for the hospitality of  the ECT* in Trento where part of this work was carried out. 

\bibliographystyle{unsrt}

\begin{thebibliography}{10}


\bibitem{Wetterich93}  C.Wetterich, Phys. Lett., {\bf B301}, 90 (1993).

\bibitem{Ellwanger93}  U.Ellwanger, Z.Phys., {\bf C58}, 619 (1993).
 
\bibitem{Tetradis94}
  N.~Tetradis and C.~Wetterich,
  Nucl.\ Phys.\ B {\bf 422},  541 (1994).


\bibitem{Morris94}  T.R.Morris, Int. J. Mod. Phys., {\bf A9}, 2411 (1994).
 
 
\bibitem{Morris94c}  T.R.Morris, Phys. Lett. {\bf B329}, 241 (1994).


\bibitem{Berges02}
J. Berges, N. Tetradis and C. Wetterich,
   Phys. Rept. {\bf 363}, 223--386 (2002).

\bibitem{Bagnuls:2000ae}
C.~Bagnuls and C.~Bervillier,
Phys.\ Rept.\  {\bf 348}, 91 (2001).


\bibitem{Canet04} L. Canet and B. Delamotte, cond-matt/0412205.
 
 \bibitem{weinberg73}
 S.~Weinberg, Phys. Rev. {\bf D8},  3497 (1973).

\bibitem{truncation} U. Ellwanger, Z. Phys., {\bf C62}, 503 (1994); 
U.~Ellwanger, M.~Hirsch and A.~Weber,
Eur.\ Phys.\ J.\ C {\bf 1} (1998) 563;
J.~M.~Pawlowski, D.~F.~Litim, S.~Nedelko and L.~von Smekal,
Phys.\ Rev.\ Lett.\  {\bf 93}, 152002 (2004);
J.~Kato,
arXiv:hep-th/0401068; C.~S.~Fischer and H.~Gies,
JHEP {\bf 0410}, 048 (2004).

\bibitem{Ellwanger94} U. Ellwanger and C. Wetterich, Nucl. Phys. {\bf B423}, 137(1994).

 
 \bibitem{alkofer} For a review, see
R. Alkofer and L. von Smekal, Phys. Rep. {\bf 353}, 281 (2001).

\bibitem{convergence} T.~R.~Morris,
  Phys.\ Lett.\ B {\bf 334}, 355 (1994).

\bibitem{club} G. Baym, J.-P.  Blaizot, M. Holzmann, F. Lalo\"e, and D.
Vautherin, Phys. Rev. Lett.  {\bf 83}, 1703 (1999).

 \bibitem{bigbec} G. Baym, J.-P.  Blaizot, M. Holzmann, F. Lalo\"e, and D.
Vautherin, Eur. Phys. J.  {\bf B24}, 107 (2001).
 
\bibitem{Andersen:2003qj}
  J.~O.~Andersen,
  Rev.\ Mod.\ Phys.\  {\bf 76}, 599 (2004).

 
   \bibitem{BigN} G.\ Baym, J-P.\ Blaizot and J. Zinn-Justin, Europhys.\
Lett.  {\bf 49}, 150 (2000).


\bibitem{latt2} P. Arnold and G. Moore,
Phys. Rev. Lett. {\bf 87}, 120401 (2001).

\bibitem{latt1} V.A.  Kashurnikov, N.~V.  Prokof'ev, and B.~V.
Svistunov, Phys. Rev. Lett. {\bf 87}, 120402 (2001).

 
 \bibitem{alpha2}
  J.~P.~Blaizot, R.~Mendez Galain and N.~Wschebor,
  \emph{Non-Perturbative Renormalisation Group
  equations and momentum dependence of $n$-point functions (II)},  hep-th/0603163.

\bibitem{Blaizot:2004qa}
  J.~P.~Blaizot, R.~Mendez Galain and N.~Wschebor,
  Europhys. Lett., {\bf 72 (5)}, 705-711 (2005).

\bibitem{PLB}
  J.~P.~Blaizot, R.~Mendez Galain and N.~Wschebor,
  Phys. Lett. {\bf B632},  571-578(2006).

\bibitem{BMWn}
J.~P.~Blaizot, R.~Mendez Galain and N.~Wschebor,
  \emph{ Non-Perturbative Renormalisation Group
  calculation of the self-energy of a scalar field},  in preparation.


\bibitem{Litim}
D.Litim, Phys. Lett. {\bf B486}, 92 (2000); Phys. Rev. {\bf D64},
105007 (2001);  Nucl. Phys. {\bf B631}, 128 (2002);
Int.J.Mod.Phys. {\bf A16}, 2081 (2001).


\bibitem{Ball95} R.D.Ball, P.E.Haagensen, J.I.Latorre and E. Moreno,
Phys. Lett., {\bf B347}, 80 (1995).

\bibitem{Comellas98} J.Comellas, Nucl. Phys., {\bf B509}, 662 (1998).

\bibitem{Canet02} L.Canet, B.Delamotte, D.Mouhanna and J.Vidal, Phys.
Rev. {\bf D67}, 065004 (2003).

\bibitem{Zinn-Justin:2002ru}
  J.~Zinn-Justin,
  \emph{Quantum field theory and critical phenomena},
  Int.\ Ser.\ Monogr.\ Phys.\  {\bf 113},  1 (2002).


\bibitem{Polchinski}
J. Polchinski, Nucl. Phys. {\bf B231}, 269 (1984).

\bibitem{Bonini}
M. Bonini, M. D'Attanasio, G. Marchesini, Nucl.Phys. {\bf B409},
441 (1993).

\bibitem{Canet03}
L. Canet, B. Delamotte, D. Mouhanna and J. Vidal,
 Phys. Rev. {\bf B68}, 064421 (2003).

\bibitem{Bervi} T.R. Morris, M.D. Turner, Nucl.Phys. {\bf B509}, 637-66 (1998);
A. Ringwald, C. Wetterich, Nucl.Phys. {\bf B334}, 506 (1990);
C. Bervillier, hep-th/0501087.


\bibitem{D'Attanasio97}
M. D'Attanasio and T.~R. Morris,
Phys. Lett. {\bf B409},  363--370 (1997).

\bibitem{Tetradis95}
N.~Tetradis and D.~F. Litim,
Nucl. Phys.  {\bf B464},  492--511 (1996).

\bibitem{Moshe:2003xn}
  M.~Moshe and J.~Zinn-Justin,
  Phys.\ Rept.\  {\bf 385}, 69 (2003)

\bibitem{refeta}
R.Guida, J.Zinn-Justin, J. Phys. {\bf A31},  8103 (1998).


\bibitem{Arnold:2000ef}
P.~Arnold and B.~Tomasik,
Phys.\ Rev.\ A {\bf 62}, 063604 (2000).


\bibitem{souza} F. de Souza Cruz, M.B. Pinto, and R.O. Ramos,
Phys. Rev. {\bf B64}, 014515 (2001); Phys.\ Rev.\ A {\bf 65},
053613 (2002).

\bibitem{Kneur04}
J.-L. Kneur, A. Neveu, M. B. Pinto, Phys. Rev. {\bf A69}, 053624 (2004).



\bibitem{Kastening:2003iu}
B.~M.~Kastening,
Phys.\ Rev.\ A {\bf 69}, 043613 (2004).
 
 
\bibitem{latt} X.~Sun, Phys. Rev. {\bf E67}, 066702 (2003).

  
  
\end{thebibliography}

\end{document}